# Decoding Polyphenol–Protein Interactions with Deep Learning: From Molecular Mechanisms to Food Applications

**Qiang Liu[1], Tiantian Wang[1], Binbin Nian[2], Feiyang Ma[1], Siqi Zhao[1], Andrés F. Vásquez[3], Liping Guo[1#], Chao Ding[1#], Mehdi D. Davari[3#]**

[1]*College of Food Science and Engineering, Nanjing University of Finance and Economics/ Collaborative Innovation Center for Modern Grain Circulation and Safety, Nanjing 210023, Jiangsu Province, PR China*

[2]*State Key Laboratory of Materials-Oriented Chemical Engineering, School of Pharmaceutical Sciences, Nanjing Tech university, Nanjing 210009, Jiangsu Province, PR China*

[3]*Department of Bioorganic Chemistry, Leibniz Institute of Plant Biochemistry, Weinberg 3, 06120 Halle, Germany*

***Corresponding authors:**

**Chao Ding**, Ph.D., Professor & **Liping Guo**, Ph.D., Lecturer
College of Food Science and Engineering
Nanjing University of Finance and Economics
Email: cding@nufe.edu.cn; guoliping07@foxmail.com

**Mehdi D. Davari,** Ph.D.
Department of Bioorganic Chemistry,
Leibniz Institute of Plant Biochemistry
Email: Mehdi.Davari@ipb-halle.de

**ABSTRACT**

Polyphenols and proteins are essential biomolecules that influence food functionality and, by extension, human health. Their interactions—hereafter referred to as *PhPIs* (polyphenol–protein interactions)— affect key processes such as nutrient bioavailability, antioxidant activity, and therapeutic efficacy. However, these interactions remain challenging due to the structural diversity of polyphenols and the dynamic nature of protein binding. Traditional experimental techniques like nuclear magnetic resonance (NMR) and mass spectrometry (MS), along with computational tools such as molecular docking and molecular dynamics (MD) have offered important insights but face constraints in scalability, throughput, and reproducibility. This review explores how deep learning (DL) is reshaping the study of PhPIs by enabling efficient prediction of binding sites, interaction affinities, and MD using high-dimensional bio- and cheminformatics data. While DL enhances prediction accuracy and reduces experimental redundancy, its effectiveness remains limited by data availability, quality, and representativeness—particularly in the context of natural products. We critically assess current DL frameworks for PhPIs analysis and outline future directions, including multimodal data integration, improved model generalizability, and development of domain-specific benchmark datasets. This synthesis offers guidance for researchers aiming to apply DL in unraveling structure–function relationships of polyphenols, accelerating discovery in nutritional science and therapeutic development.

## 1. Introduction

### 1. Polyphenol-protein interactions (PhPIs)

Polyphenols are broadly defined as complex compounds containing one or more phenolic hydroxyl groups and are widely found in various plants like fruits, vegetables, legumes, cereals, and tea (Jiang et al., 2024). With a diverse chemical structure, polyphenols have been shown to possess antioxidants, antibacterial, and anti-inflammatory properties (Sun et al., 2024). They can also lower blood glucose levels, reduce the risk of cardiovascular diseases, and improve cholesterol levels (Wan et al., 2024). Due to the inability of the human body to synthesize them directly and their irreplaceability by other nutrients, they have garnered significant attention in fields such as medicine, food science, nutrition, and health care (Álvarez-Carretero et al., 2018; Alves et al., 2023)

Polyphenols are widely recognized for their ability to bind to proteins (PhPIs)—the primary focus of this work—but they can also interact with other biomolecules such as DNA, lipids, and polysaccharides (Amoako and Awika, 2016). Among these interactions, PhPIs are particularly significant due to its implications in enzyme modulation, signaling pathways, and nutrient bioavailability (Cory et al., 2018; Jakobek, 2015; Kardum and Glibetic, 2018; Liu, J. et al., 2024).

Polyphenols interact with proteins through both non-covalent mechanisms—such as hydrogen bonding, hydrophobic interactions, and electrostatic forces—and covalent mechanisms, including the formation of C–S and C–N bonds or linkage with sulfhydryl (–SH) groups on proteins (Rashid et al., 2022).

PhPIs are observed in various protein types, including plant-derived proteins such as soy protein isolate (SPI), whey protein, gelatin, and specific enzymes. For instance, the non-covalent binding between Lycium barbarum polysaccharide (LBP) and SPI enhances gel strength, thermal stability, and β-sheet content (Xing et al., 2024). Yan et al. (2020) investigated camphor protein isolates and observed positive effects of polyphenols on both physicochemical properties and functional characteristics of the protein isolates. Similarly, when soybean proteins bind to anthocyanins, changes in functional characteristics and structural alterations are also observed (Wang, Yan et al., 2022; Yan et al., 2020). Both non-covalent/covalent PhPIs can induce the formation of highly ordered or disordered structures (Meng and Li, 2021), which not only influences the conformation and stability of proteins (Dai et al., 2019) but also impacts its biological activity and physiological effects such as enhancing antioxidant (Liu, X. et al., 2021), antimicrobial (Li et al., 2024), anti-glycation capacity (Ravichandran et al., 2021), as well as affecting digestibility (Günal-Köroğlu et al., 2023; Ozdal et al., 2013) and solubility (Norn et al., 2021). Therefore, exploring the mechanisms underlying PhPIs holds immense theoretical and practical significance.

Polyphenols interact with a wide spectrum of protein targets, influencing both nutritional and pharmacological outcomes. PhPIs span enzymatic and non-enzymatic proteins, often modulating key physiological pathways. Among enzymes, polyphenols prominently inhibit digestive hydrolases, especially α-amylase and α-glucosidase, thereby reducing carbohydrate digestion and glucose absorption (Corkovic et al., 2022; Visvanathan et al., 2024). They also interact with oxidoreductases like tyrosinase, implicated in melanin biosynthesis and pigmentation disorders (Fan et al., 2019; Yu et al., 2019), and transferases, such as certain kinases, modulating phosphorylation-dependent signaling cascades (Islam et al., 2024; Kang et al., 2011). Beyond enzymatic proteins, polyphenols bind to a diverse range of non-enzymatic targets, including salivary proline-rich proteins, structural proteins like casein and actin, signaling molecules such as MAPKs and NF-κB components, and carrier proteins like serum albumin (Morzel et al., 2022; Sonawane and Chinnathambi, 2021; Wang, X. et al., 2024; Zhang, Y. et al., 2024). Such examples will be examined in more detail throughout the following sections, according to their biological roles. Altogether, this multifaceted interaction landscape underscores the pleiotropic nature of polyphenols and their capacity to influence health at multiple molecular levels—from digestion and metabolism to cellular signaling and structural integrity.

### 1.2. Challenges in understanding PhPIs

In recent years, numerous scholars have conducted extensive research on the mechanism behind PhPIs and have clarified the ways in which they interact **(Fig.1)**. Polyphenols primarily engage with proteins through covalent and non-covalent bonds (including hydrophobic interactions, hydrogen bonding, electrostatic interactions, and van der Waals forces), resulting in the formation of complexes that significantly impact protein structure, function (Masoumi et al., 2024). To reveal PhPIs and their conformational changes upon binding, researchers have used various techniques (Perez-Gregorio and Simal-Gandara, 2017), including fluorescence spectroscopy, circular dichroism (CD) spectroscopy, dynamic light scattering (DLS), Fourier transform infrared (FT-IR) spectroscopy, isothermal titration calorimetry (ITC), and nuclear magnetic resonance (NMR). Additionally, molecular docking is widely utilized to efficiently analyze the biological and biochemical roles of proteins, subsequent changes in ligand binding, and to predict the structures and activities of binding targets (Shahidi and Dissanayaka, 2023).

Gong and coworkers recently investigated the changes in the secondary structure of casein and whey proteins upon binding to anthocyanins by spectroscopy techniques such as (FTIR) and CD, and found that the α-helix content was significantly reduced, while the β-folding and random curling content were increased (Gong et al., 2021). Thermodynamic analysis showed that the reaction was a spontaneous process mainly driven by hydrogen bonding and van der Waals forces, suggesting that the protective effect of proteins on anthocyanins may contribute to their stability. Meanwhile, the fluorescence intensity of casein and whey proteins decreased significantly with increasing anthocyanin concentration, and the fluorescence peaks were red-shifted, further suggesting that the microenvironment of the proteins changed (Geng et al., 2023). In addition, molecular docking studies showed that α-casein was able to bind to mallowin-3-O-galactopyranoside (M3G) via hydrogen bonding, and the hydroxyl group and galactose of M3G interacted with amino acid residues of the protein.

However, given the diversity of polyphenol and protein structures and the differences in their behavior in different environments, there are still many challenges in studying PhPIs. The diversity of polyphenol and protein structures and species, differences in the mode of interaction, the influence of environmental factors, and the limitations of measurement techniques and methodologies may affect how and to what extent they interact (Xue et al., 2025). These factors can significantly affect the binding sites of polyphenols and proteins, the degree of binding, and the structure of the complexes, which in turn can alter the functional and nutritional properties of the complexes (Poklar Ulrih, 2017). In addition, the traditional experimental characterization methods mentioned above are problematic in terms of data analysis, computational time and cost, as well as accuracy and reliability (Ozdal et al., 2013).

Although computational approaches provide new insights, they still face the challenges of model dependence, long computation times, and adaptability to real biological systems. MD simulations, for example, based on physical free energy calculations, demand high computational resources in the study of polyphenol-protein binding modes. The predictive accuracy typically deviates from experimental values by approximately ±1.0 kcal/mol (da Silva et al., 2024) Thus, the structural diversity of polyphenols and proteins and the differences in their behavior in different environments further increase the need for large-scale datasets and complex models.

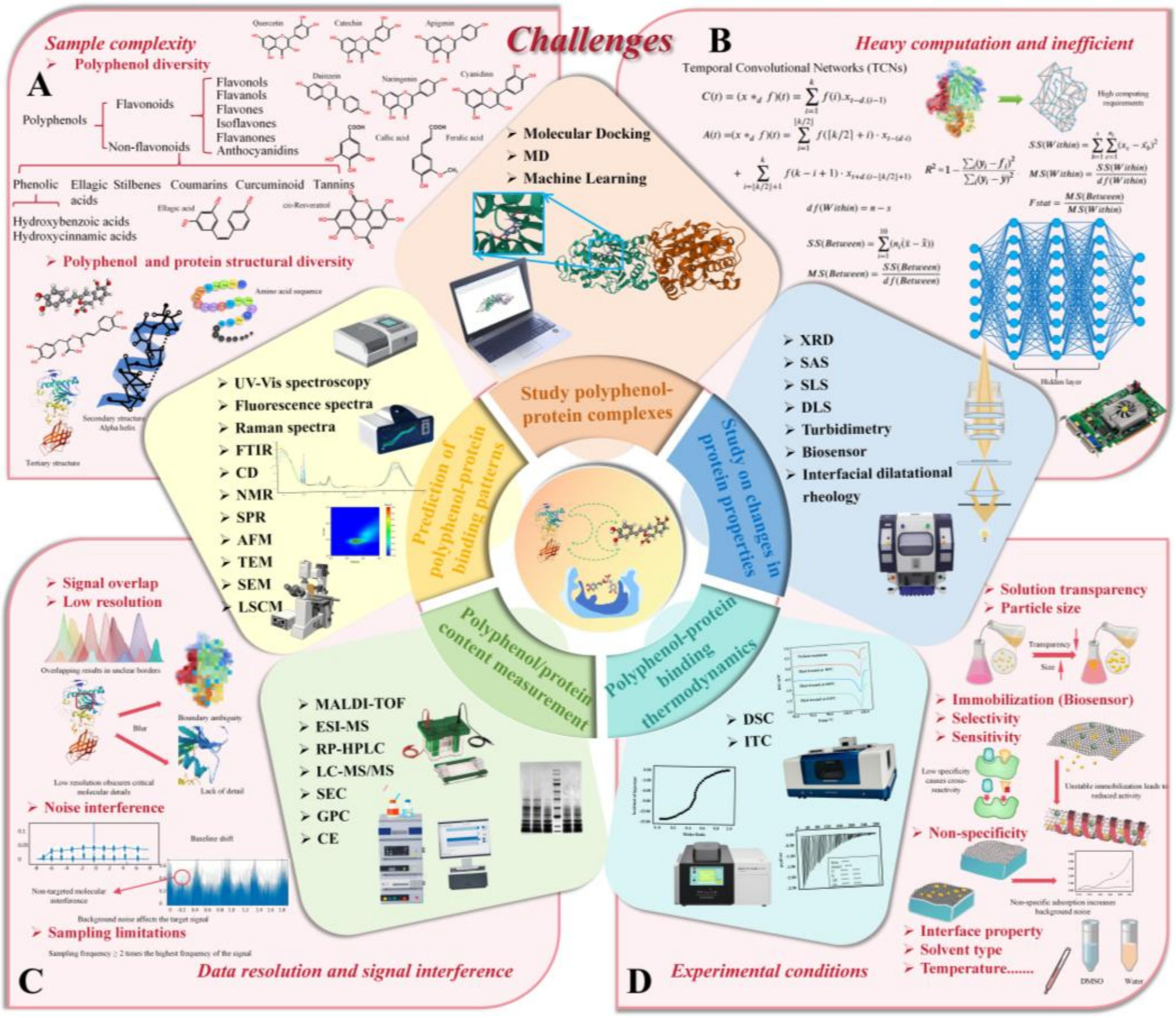

**Figure 1: Approaches and challenges in studying PhPIs.** (**A**) Sample complexity: The diverse types and structures of polyphenols and proteins complicate sample pretreatment and characterization, making the analysis of intermolecular interactions especially challenging in spectroscopy, light scattering, and mass spectrometry. (**B**) High Computational Load and Low Efficiency: Techniques such as molecular docking and MD predict interaction patterns but are inefficient at capturing dynamic processes. Additionally, methods like sub-docking and ML (machine learning) (e.g., temporal convolutional networks) require complex parameter tuning and iterative modeling, which further limits efficiency. (**C**) Data Resolution & Signal Interference: Low resolution, overlapping signals, and non-target noise in spectroscopy and microscopy hinder accurate analysis of microscopic interactions. (**D**) Experimental Conditions: Variations in solvent properties, particle sizes, and sensor immobilization further challenge reproducibility and data fidelity.

Illustration: AFM (Atomic force microscope), SEM (Scanning electron microscope), TEM (Transmission Electron Microscope), LSCM (Laser scanning confocal microscope), SEC (Size-exclusion chromatography), MALDI-TOF (Matrix-Assisted Laser Desorption/Ionization Time-of-Flight Mass Spectrometry), ESI-MS (Mass spectrometry Electrospray ionization mass spectrometry ), CE (Capillary Electrophoresis), RP-HPLC (Reverse-phase high-performance liquid chromatography), LC-MS/MS (Ultrafiltration combined with liquid chromatography tandem mass spectrometry), SEC (Size-exclusion chromatography), GPC (Gel Permeation Chromatography), XRD (X-ray diffraction), SAS (Small-Angle Scattering), SLS (Static light scattering).

### 1.3. Implications of DL algorithms for solving interaction relationship understanding

Artificial Intelligence (AI) is a field of science and technology that studies how to enable computers to simulate and perform human intelligence tasks. It is dedicated to developing intelligent systems that can sense, understand, learn, reason, make decisions, and interact with humans. The background of AI dates back to the 1950s, when initial AI research focused on rule-based reasoning and the development of expert systems (Haenlein and Kaplan, 2019). With the advancement of computer technology and algorithms, especially the rise of ML and DL, AI has begun to usher in explosive development.

DL algorithms are a type of ML method based on artificial neural networks, known for their powerful nonlinear modeling capabilities and adaptability to large-scale data. In recent years, DL algorithms have made a series of breakthroughs in biology, encompassing protein design (Dauparas et al., 2022; Ding et al., 2022; Norn et al., 2021), structure prediction (Baek et al., 2021; Huang, J. et al., 2023; Yang, Z. et al., 2023), protein function prediction (Giri et al., 2021; Kulmanov et al., 2021; Zhao et al., 2024), protein-protein interaction (PPI) prediction (Kaundal et al., 2022; Lei et al., 2021; Zheng, J. et al., 2023), protein-ligand interactions (Cai et al., 2024; Moon et al., 2024b) and drug discovery (Allenspach et al., 2024; Atz et al., 2024; Tropsha et al., 2024). A notable example is AlphaFold, which has made significant progress in predicting the three-dimensional structure of proteins from amino acid sequences, addressing a challenge that has persisted for over fifty years and achieving accuracy comparable to experimentally determined structures (Jumper et al., 2021).

DL has also been applied to various aspects of protein-ligand docking, from pose prediction to virtual screening activities (Crampon et al., 2022). Publications in the scientific literature have shown that DL-based strategies enable faster docking schemes, more accurate scoring functions and virtual screening of libraries containing billions of compounds (Gentile et al., 2022). A DL model proposed by (Stärk et al., 2022) called EquiBind directly predicts the structure of protein-ligand complexes without relying on population-based optimization methods. As a result, its predictions are significantly faster than other methods, taking less than a second per prediction. Likewise, the DL classifier DFCNN, introduced by (Zhang, H. et al., 2022), enabled the discovery of five novel compounds with activity against the validated target trypsin I. The model uses only sequence information from ligands and proteins. Because the model uses only ligand and protein sequence information, each prediction takes only a few seconds, representing a speedup of approximately four to five orders of magnitude compared to AutoDock Vina. In addition, DeepDocking can reduce the number of docked molecules required to screen very large docking databases by up to 50 times compared to traditional molecular docking methods, while losing only about 10% of the virtual hit rate (Gentile et al., 2020).

The utilization of DL algorithms to assist in elucidating PhPIs has emerged as a novel research direction. The wealth of protein data serves as a solid foundation for data-driven hypothesis generation and biological knowledge discovery. With the continuous advancement of AI, DL can automatically extract nonlinear, intrinsic, abstract and complex models from large-scale data without a prior knowledge (Shi et al., 2021). Given that conformational changes in PhPIs can induce folding or unfolding of protein molecules, thereby altering their structure and properties and significantly affecting their functional properties and biological activities, research in this area can benefit. Indeed, incorporating Deep Learning (DL) algorithms into such research, while posing challenges, proves highly effective (Jiang et al., 2019). Tools such as molecular docking and kinetic simulation using DL enable comprehensive screening of competitive or non-competitive PhPIs with respect to site specificity or nonspecificity, which can help to understand the interactions at the molecular level (Shahidi and Dissanayaka, 2023). Therefore, the integration of DL algorithms with experimental techniques holds significant importance for a better insight into PhPIs. This integration can provide a scientific basis for fully harnessing the functions and properties of polyphenol-protein complexes or conjugates, while also broadening their application areas through directed evolution.

## 2. Basic Properties of Polyphenols, Proteins, and PhPIs

To explore the PhPIs, DL algorithms must first master the basic properties and functional characteristics of polyphenols and proteins, which is the basis for establishing prediction and training models, designing experiments and understanding their biological effects. To screen out effective information that may be neglected through big data processing and data mining methods and then assist researchers to further study the interaction and influence between polyphenol and protein.

### 2.1. Classification and characterization of polyphenols

Polyphenols represent a group of secondary metabolites derived from plants that are widely found in vegetables, fruits, and grains (Okumura, 2021). They are characterized by the presence of at least two aromatic rings with one or more hydroxyl substituents (Biluca et al., 2020). Based on their chemical structure and source, polyphenols can be categorized into flavonoids (such as flavones, flavonols, flavanones, flavan-3-ols, isoflavones, anthocyanins) and non-flavonoids (including phenolic acid, stilbene phenol, coumarin, curcumin and lignans) (Li, Y. et al., 2021), as shown in **Figure 2**.

Polyphenols are recognized as natural antioxidants, primarily attributed to their ability to inhibit reactive oxygen radicals (Huang, W. et al., 2023) and hydrogen peroxide, and the oxidative damage to DNA and suppress lipid peroxidation also can be mitigated. Additionally phenolics can regulate the content of oxidized malondialdehyde (MDA) (Chen et al., 2023) and reduce the level of 8-hydroxydeoxyguanosine (8-OHdG) by enhancing the activity of various antioxidant enzymes in the body, such as glutathione peroxidase (Feng et al., 2023) and superoxide dismutase (López-Huertas and Palma, 2020), which in turn can inhibit reactive oxygen species generation, scavenge free radicals and peroxides, and exert their antioxidant effects to protect DNA from oxidative damage (Liu and Zheng, 2002). Besides their antioxidant activity, polyphenols exhibit notable anti-inflammatory properties. They mitigate inflammatory by disrupting the synthesis and release of mediators such as tumor necrosis factor-α (TNF-α) and interleukin-1β (IL-1β). Moreover, they also modulate inflammation-related signaling pathways, including nuclear factor κB (NF-κB) (Baron et al., 2021) and MAPK pathways (Behl et al., 2022), which is crucial for preventing and treating inflammatory diseases. Additionally, the anti-infective properties of polyphenols (Alshehri et al., 2022; Cai et al., 2023; Jakobek and Blesso, 2024; Sharifi-Rad et al., 2023) are achieved by interfering with the growth and reproduction of pathogenic microorganisms, including bacteria, viruses, fungi, and other pathogens. At the same time, they can enhance the immune system, promote the activation and antimicrobial action of immune cells accelerating the clearance of infections.

In addition to the above properties, biological abilities such as the reduction of cardiovascular disease risk (Huang, Y. et al., 2024), antiviral (Montenegro-Landívar et al., 2021), antibacterial (Cai et al., 2019; Gómez-Maldonado et al., 2020; Pandey et al., 2023;

Rampone et al., 2021; Sun et al., 2021; Tian et al., 2021; Xie et al., 2021; Yan et al., 2021; Yang, R. et al., 2023; Yemiş et al., 2022; Zhao et al., 2022), anti-allergic (Kojima et al., 2000; Zeng et al., 2023), and immune-enhancing (Thirumdas et al., 2021) properties are possessed by polyphenols. They are of concern in both natural medicine and functional food research.

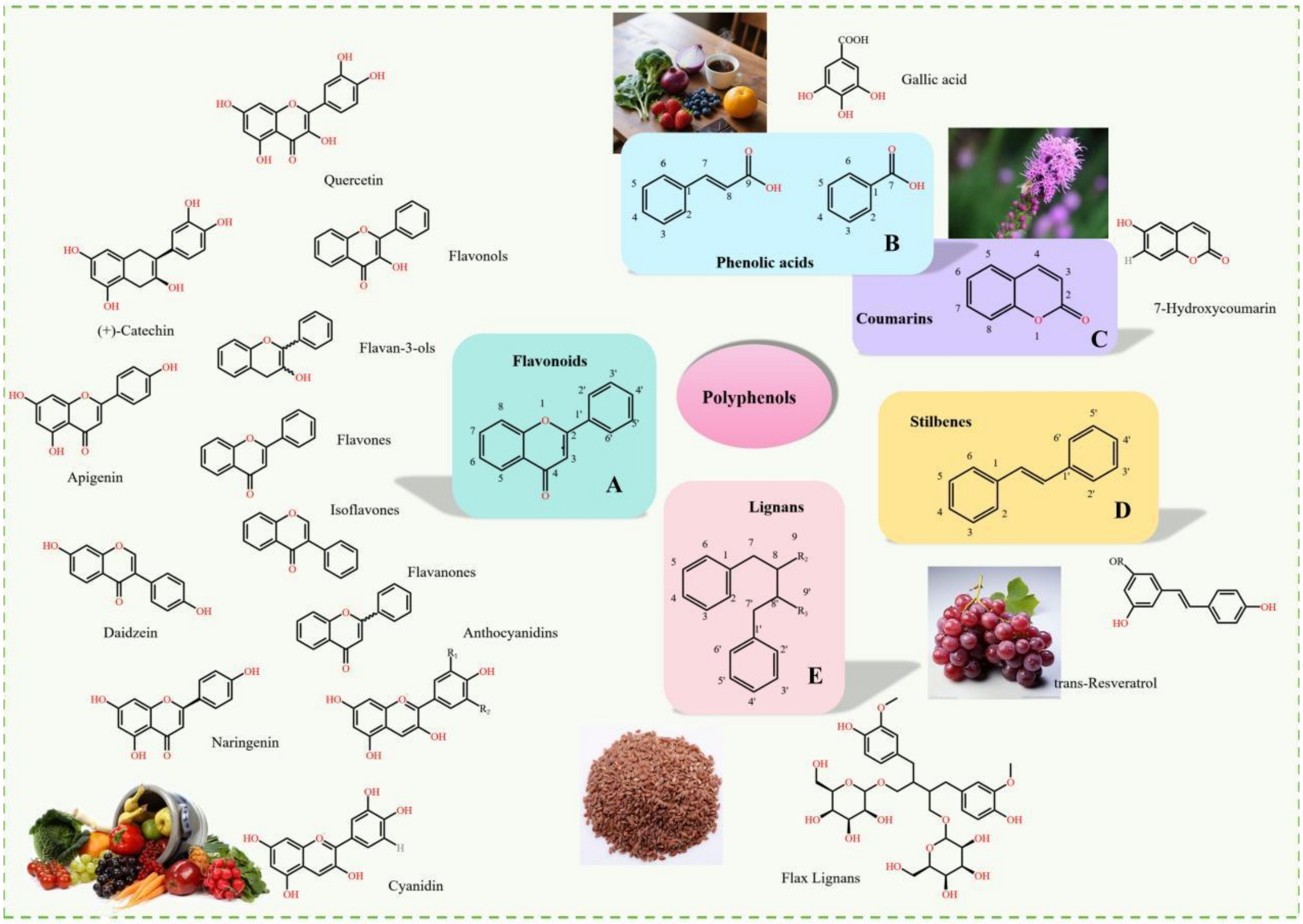

**Figure 2: Classification of polyphenols and the chemical formulas of representative compounds.** Polyphenols are plant secondary metabolites characterized by the presence of one or more phenolic hydroxyl groups. They are broadly classified into five major groups based on their structural features: Flavonoids (**A**), Phenolic acids (**B**), Coumarins (**C**), Stilbenes (**D**), and Lignans (**E**). This figure illustrates the core structures of each group along with representative compounds, such as quercetin (flavonoids), gallic acid (phenolic acids), 7-hydroxycoumarin (coumarins), trans-resveratrol (stilbenes), and secoisolariciresinol diglucoside (lignans), highlighting the structural diversity and biological relevance of dietary polyphenols.

**2.2. Classification and function of proteins**

Polyphenols constitute a structurally diverse class of plant-derived compounds that exhibit broad protein-binding capabilities. The PhPIs with different protein types are dictated by the molecular architecture of both the polyphenols and the target proteins, resulting in protein-specific structural and functional outcomes.

Within enzyme proteins, polyphenols predominantly modulate catalytic activity and conformational stability through non-covalent interactions—such as hydrogen bonding and hydrophobic contacts—at or near active sites (Liu and Zheng, 2002). Additionally, oxidative conversion of polyphenols to o-quinones enables covalent bonding with enzyme side chains, further influencing enzymatic function (Le Bourvellec, C. and Renard, C.M., 2012). Flavonoids including apigenin and quercetin exemplify this phenomenon by exerting competitive or non-competitive inhibition on pancreatic lipase, effects closely tied to critical structural motifs such as the C2=C3 double bond (Li, M.-M. et al., 2023). Beyond inhibitory activity, polyphenol binding induces conformational alterations, enhancing thermal stability and decelerating enzyme denaturation (Li, S. et al., 2021), underscoring a complex regulatory mechanism.

Structural proteins—collagens, caseins, and myofibrillar proteins—play pivotal roles in maintaining extracellular matrix integrity and tissue mechanical properties. Polyphenols facilitate both covalent and non-covalent cross-linking among protein molecules, markedly improving thermal resilience and proteolytic resistance (Le Bourvellec, C. and Renard, C.M., 2012). For instance, proanthocyanidins form covalent cross-links within dentin collagen, significantly enhancing its biostability (Hass et al., 2021). Similarly, quercetin and gallic acid undergo laccase-catalyzed oxidation to covalently bind acid-soluble collagen, stabilizing the protein, improving emulsification properties, and reducing digestibility (Wang, Y. et al., 2023). In dairy systems, PhPIs with caseins modulate protein solubility and emulsion stability, culminating in improved textural and sensory qualities (van de Langerijt et al., 2024). Furthermore, polyphenols engage myofibrillar proteins via hydrogen bonding and hydrophobic forces to modulate secondary and tertiary structures, thereby enhancing gel density and water retention, attributes essential for food quality (Jiang et al., 2025).

Transport and storage proteins such as serum albumin and transferrin perform crucial roles in molecular transport and metabolic regulation (Fanali et al., 2012; Papanikolaou and Pantopoulos, 2017). As small-molecule ligands, polyphenols selectively bind specific ligand sites on these proteins, modulating their affinity for drugs or endogenous molecules, thereby impacting pharmacokinetics and bioavailability (Fu et al., 2025). Notably, epigallocatechin gallate (EGCG) forms stable complexes with serum albumin, bolstering structural stability and antioxidant capacity (Li, Z. et al., 2022). Polyphenol-protein complexes also exhibit favorable biocompatibility and stability, demonstrating superior drug-loading and controlled-release characteristics in delivery systems (Ramesh and Mandal, 2019). Recent advances have exploited polyphenol-mediated protein assembly to engineer nanoscale carriers, enabling precise targeted drug delivery and highlighting the expansive potential of polyphenols in biomaterials science (Zhang, H. et al., 2024).

Signal and receptor proteins constitute key targets mediating the bioactivity of polyphenols. These compounds bind specific receptor sites, triggering conformational shifts that regulate downstream signaling pathways involved in cell proliferation, apoptosis, and inflammatory responses (Byun et al., 2012). For example, polyphenols such as curcumin and quercetin inhibit aberrant activation of the JAK-STAT pathway, contributing to anti-inflammatory and anticancer effects (Zalpoor et al., 2022). Additionally, polyphenols activate AMPK signaling, enhancing metabolic regulation and offering therapeutic promise for metabolic diseases such as diabetes (Kim et al., 2009).

### 2.3. Mechanisms of PhPIs

Polyphenols and proteins are crucial in biological processes, and the mechanisms underlying PhPIs are of great interest in food science. The system's heat is affected by changes in the strength of molecular interactions when PhPIs occur. Information about the combined thermodynamic parameters ($\Delta H$, $\Delta S$, and $\Delta G$) has been employed to ascertain the characteristic of the forces involved (Li, Y. et al., 2021). A substantial body of research has demonstrated that PhPIs mechanisms can be classified into two main categories: non-covalent and covalent.

Non-covalent interactions—including hydrogen bonds, hydrophobic interactions, electrostatic interactions, and van der Waals forces—play a central role in mediating PhPIs (**Fig. 3A**). While generally weaker and reversible compared to covalent bonds, these interactions are fundamental to binding affinity and specificity. Among these, hydrogen bonds often play a dominant role, as polyphenols contain multiple hydroxyl groups capable of acting as both hydrogen bond donors and acceptors (Shahidi and Dissanayaka, 2023). Their aromatic structure further contributes to directional and stabilizing polar interactions with protein residues (Nassarawa et al., 2023). Ferulic acid (FA), quercetin (QT), and vanillic acid (VA) can interact with β-lactoglobulin through hydrogen bonds (Zhang, S. et al., 2022). Similarly, hydrogen bonding between soy protein and catechin/epigallocatechin gallate (EGCG) in tea polyphenols has been explored (Dai et al., 2024). The interaction between EGCG and human serum albumin is also dependent on hydrogen bonds and involves multiple binding sites, including Glu141, Leu182, Tyr138, and Tyr161 (Chanphai and Tajmir-Riahi, 2019). In addition to these polar interactions, polyphenols may also engage in hydrophobic interactions, primarily involving their nonpolar aromatic rings and the nonpolar regions of proteins (Baba et al., 2021). For instance, hydrophobic interactions at specific amino acid residues facilitate the engagement of β-lactoglobulin with theaflavins (Xu et al., 2019). However, the strength and significance of hydrophobic interactions are often context-dependent, influenced by the structural environment of the binding site and solvent exposure (Shahidi and Dissanayaka, 2023). Furthermore, van der Waals forces, though individually weak, contribute cumulatively by stabilizing close-range contacts between the ligand and protein (Xiao et al., 2023). For instance, the binding of anthocyanin-3-rutinoside (C3R) and epicatechin (EC) from mulberry polyphenols to β-lactoglobulin (β-LG) involves both hydrogen bonding and van der Waals forces (Yuan et al., 2024). In another study, the cross-linking of kaempferol and QT with microbial glutamine transferase (TGase) at temperatures between 293 and 313 K was found to be primarily mediated by hydrogen bonding and van der Waals forces (Zhang, Y.J. et al., 2021). Finally, electrostatic interactions are generally considered minor contributors to polyphenol–protein binding. This is because polyphenols are typically not charged at physiological pH or in most food matrices, which limits their ability to engage in electrostatic interactions with proteins, except for certain ionizable phenolic acids (Ebrahimi et al., 2025; Shahidi and Dissanayaka, 2023). Moreover, the high dielectric constant of aqueous environments further diminishes the strength of long-range ionic interactions (Collins, 2012). It has been demonstrated that phenolic acids with low pKa values (e.g., cinnamic acid derivatives and FA) interact under neutral conditions with highly electronegative polyphenol hydroxyl groups to deprotonate positively charged protein such as the ε-amino of lysine (Li, Y. et al., 2021).

On the other hand, covalent PhPIs are typically irreversible. This process primarily involves the oxidation of polyphenols to produce semi-quinone radicals or quinones, which subsequently form Schiff bases or Michael adducts due to their electrophilicity interaction with nucleophilic amino acid residues (**Fig. 3B**) (Yang, C. et al., 2019). the presence of caffeic acid cysteine (CA-Cys) and chlorogenic acid cysteine (CGA-Cys) bound to protein in milk samples treated with caffeic acid (CA) and chlorogenic acid (CGA) was detected by (Poojary et al., 2023), thereby confirming the covalent bonding between milk protein (MP) and CA/CGA. Furthermore, phenolics can also interact with sulfhydryl, amino, guanidino, or imidazole groups on proteins or peptides, where free sulfhydryl groups have been identified as more likely to covalently cross-link than other groups (Pham et al., 2019). Waqar and colleagues (Waqar et al., 2022) investigated the covalent modification of 4-methylbenzoquinone (4MBQ) with β-LG in bovine milk juice and found that compared with β-LG, the loss of free thiols and amines in β-LG after covalent modification was significant. The specific modifications of 4MBQ were identified at Cys, Lys, Arg, His, and Trp of β-LG (β-LQ).

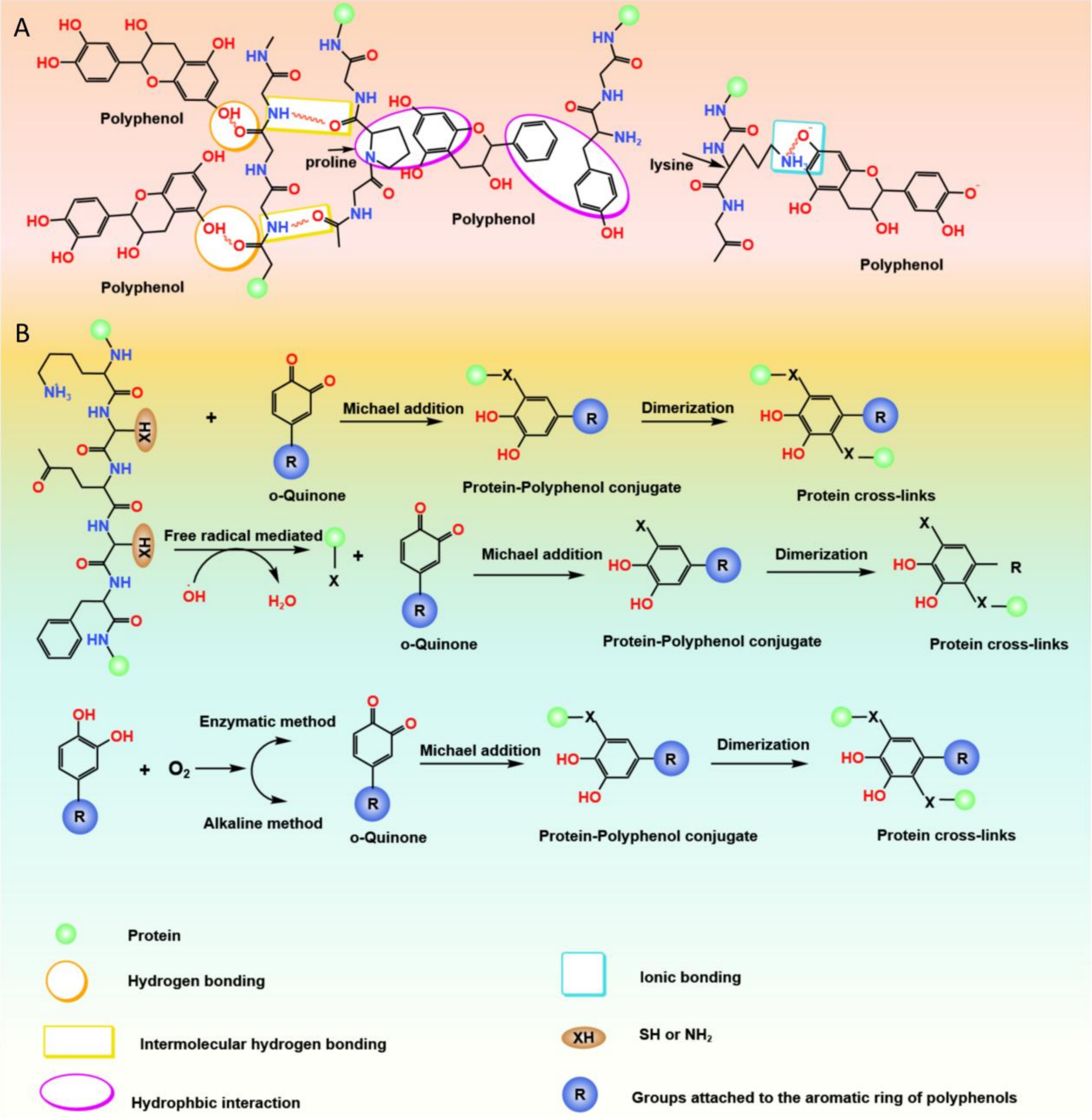

**Figure 3: Non-covalent and covalent PhPIs**. (**A**) Non-covalent PhPIs. Polyphenols interact with proteins through various non-covalent interactions, including hydrogen bonds, intermolecular hydrogen bonds, hydrophobic interactions, and ionic bonds. Hydrogen bonds (orange circles) are formed between the phenolic hydroxyl groups of polyphenols and hydrophilic amino acid residues in proteins, such as lysine and proline. Intermolecular hydrogen bonds (yellow rectangles) occur between polyphenol molecules, promoting self-aggregation and enhancing the stability of polyphenol-protein complexes. Hydrophobic interactions (purple ovals) arise between the aromatic rings of polyphenols and the nonpolar regions of hydrophobic amino acid residues, such as proline and phenylalanine, in proteins. Ionic bonds (blue boxes) are established under alkaline conditions, where deprotonated hydroxyl groups of polyphenols interact with positively charged protein side chains, such as lysine amino groups. (Adapted from (Nassarawa et al., 2023) (**B**) Covalent PhPIs. The covalent cross-linking mechanism between polyphenols and proteins involves several key processes. Polyphenols are oxidized to reactive ortho-quinones under enzymatic or alkaline conditions. Enzymatic oxidation occurs with oxygen via polyphenol oxidase or peroxidase, while non-enzymatic oxidation happens in an alkaline environment. The resulting ortho-quinones react with nucleophilic groups in proteins, such as thiol (-SH) or amino (-NH2) groups, through Michael addition reactions, forming stable polyphenol-protein conjugates that enhance the stability of the complex. Additionally, ortho-quinones can dimerize with other polyphenols or conjugates to create highly cross-linked protein networks, improving mechanical strength and antioxidant properties. Under free radical conditions, polyphenols can also form covalent bonds with proteins through polyphenol radicals; this reaction establishes cross-linking structures common in environments induced by oxidative stress or radiation. (Adapted from (Feng et al., 2023). Illustration: Green circle = protein, orange circle = hydrogen bond, yellow rectangle = intermolecular hydrogen bond, purple oval = hydrophobic interaction, blue box = ionic bond, brown "XH" = thiol (-SH) or amino ($-NH_2$), "R" = polyphenol functional groups.

PhPIs profoundly impact the structural and functional attributes of proteins and polyphenols. Pi and coworkers (Pi et al., 2023) studied the covalent interaction of SPI with gallic acid (GA), CA, and tannic acid (TA). The findings indicated that the α-helical of proteins in SPI-polyphenol conjugate disrupted, while surface hydrophobicity and free sulfhydryl groups were enhanced. Furthermore, the antioxidant capacity and emulsification performance of the conjugate were also improved. Similarly, Guo and collaborators (Guo et al., 2024) showed that incorporation of CA, GA, epicatechin gallate (ECG), and EGCG led to changes in the secondary and tertiary structure of MP, among which GA can effectively improve the gel properties of MP. The emulsification performance of perilla seed powder protein can be also improved by the addition of gallic acid (GA), protocatechuic acid (PCA) and caffeic acid (CA) (Wang, D. et al., 2024). Moreover, polyphenols can bind to enzymes such as α-amylase and alter their activity. It was discovered that the inhibition of α-amylase activity was facilitated by TA following the formation of ternary complexes with bovine serum albumin (BSA) and whey protein (IWP) (Li, W. et al., 2023). Several investigations have explored the impact of PhPIs on protein bioavailability. In vitro digestion experiments involving puerarin (PUE) with zein and SPI, forming zein-PUE and SPI-PUE complexes revealed that protein digestion was promoted by PUE, resulting in a 1.15% and 2.11% increase in the digestibility of zein-PUE and SPI-PUE, respectively. This indicates that protein digestion characteristics can be enhanced by PUE through its influence on their physicochemical properties and spatial structure (Pan et al., 2023). In addition, polyphenols have beneficial effects in the treatment of type 2 diabetes and metabolic syndrome through various mechanisms (Lacroix et al., 2018). For example, the green tea polyphenol epigallocatechin-3-gallate inhibits gluconeogenesis by activating 5'-AMP-activated protein kinase (AMPK) via $Ca^{2+}$/calmodulin-dependent protein kinase kinase (CaMKK) (Shahwan et al., 2022). Resveratrol has also been shown to stimulate AMPK through the NAD-dependent deacetylase sirtuin-1 (SIRT1), thereby improving mitochondrial function in vivo (Ungurianu et al., 2023).

Despite the extensive literature on the mechanism behind PhPIs, it remains challenging to ascertain their contribution to the binding process due to the energy exchanges and interactions that occur simultaneously between proteins and polyphenolic ligands (Ji et al., 2020). The binding sites, degree of binding, and structural configuration of the resultant protein-polyphenol complexes may be influenced by the structural diversity of proteins and polyphenols, variable interaction modes, environmental factors such as pH and temperature, and constraints imposed by measurement techniques (Xue et al., 2025). Through Fourier transform infrared spectroscopy analysis by Ramos-Pineda, the binding stability of the interaction between ferulic acid (FA) and zein was stronger under alkaline conditions (pH = 9) than under acidic and neutral conditions (Wang, Q. et al., 2022). Substantial challenges persist in fully elucidating PhPIs.

## 3. Methods for studying PhPIs

### 3.1. Traditional methods used to study PhPIs

Recent studies on PhPIs have attracted much attention. **Table 1** outlines the parameters and methods utilized. However, given their complexity, no single technique can provide comprehensive insights. Consequently, a combination of multiple techniques is often employed for characterization. In traditional studies, a range of analytical techniques is applied by researchers, including spectroscopic techniques (El-Messery et al., 2020) such as UV-VIS, fluorescence and infrared spectroscopy, thermodynamic methods (Geng et al., 2023) like isothermal titration calorimetry, and microscopic methods (Lv et al., 2022), among others. In UV-VIS spectroscopy, PhPIs are inferred by observing absorption peaks. Information about the fluorescence intensity of proteins labeled with fluorophores and bound to the polyphenol is provided by fluorescence spectroscopy. Infrared spectroscopy can be employed to observe the observation of vibrational modes between molecules, thereby shedding light on the nature of their interactions (Czubinski and Dwiecki, 2017).

The advent of sophisticated technology and instrumentation has led to the growing application of spectroscopy-based techniques in research within this field, including nuclear magnetic resonance (NMR) spectroscopy (Ferrer-Gallego et al., 2017), light scattering, mass spectrometry (MS), and chromatography (CS). NMR spectroscopy can be employed to obtain information regarding protein aggregation, dynamics and stability. A rapid assessment of kinetic parameters and particle size distribution of protein-polyphenol complexes is offered by dynamic light scattering (DLS) (Ren et al., 2022). And the structure of biomolecular assemblies with low partitioning rates can be investigated using X-ray small-angle scattering (XLS) (Chen, D. et al., 2024), which provides details on their shape, size, internal structure, and organization, as well as kinetic information on molecular folding and assembly processes. Another valuable tool for studying biological monomer structures and providing kinetic data on molecular folding and assembly is High-performance affinity chromatography (HPAC) (Cao et al., 2019). Additionally, HPAC can be used to assess the binding rates of diverse polyphenol structures to proteins. Liquid chromatography-tandem mass spectrometry (LC-MS/MS) (Sun et al., 2022) is employed to characterize binding sites and chemical structures of relevant adducts.

Furthermore, molecular docking and MD simulation are widely used in drug screening and structure-based drug design. Recently, they have been applied to PhPIs (Pinzi and Rastelli, 2019; Shahidi and Dissanayaka, 2023) to predict binding sites, estimate affinities, and assess complex stability and dynamics. Molecular docking primarily simulates the theoretical feasibility and action of interactions, providing results such as binding free energy, binding site, binding conformation, and interaction forces between proteins and polyphenols. A study conducted recently by (Wang, T. et al., 2023), involved molecular docking of SPI and TA. The findings indicated that TA tended to bind in the hydrophobic cavity of the protein by selecting the optimal scoring location for analysis. Likewise, MD simulations are used to simulate the dynamic evolution of polyphenol-protein complexes over time scales based on atomic force fields. Through this approach, biomolecule trajectories are determined using Newtonian equations of motion for all atoms in the system (Hollingsworth and Dror, 2018). Detailed information on binding stability, dynamic conformation, and binding sites is yielded by the simulations, thereby elucidating the complexity of PhPIs. (Zhang et al., 2023) utilized MD simulations to calculate the center-of-mass distance (DCOM) between FA, QT, VA, and β-LG. It was revealed that the DCOM between FA and β-LG decreased from 1.4 nm to nearly 1.0 nm during dynamic simulations within 100 ns, suggesting that FA gradually extends into the hydrophobic pocket of β-LG, forming a more stable FA-β-LG compound. Furthermore, MD is also a representative technique for the identification and analysis of tunnels, which can enhance the catalytic function of lipase by simulating ligand transport processes, including activity, substrate specificity, and stability (Chong et al., 2024).

In elucidating the structural principles of PhPIs, Quantitative Structure-Activity Relationship (QSAR) modeling also plays a significant role. As a classic and fundamental computational approach, QSAR modeling is widely used to predict the biological activity of compounds, particularly natural polyphenolic substances. Traditional QSAR techniques rely on extracting molecular descriptors—such as hydrophobicity, molecular weight, polar surface area, and functional group counts—and applying statistical or ML algorithms like multiple linear regression (MLR), SVM, and RF to model the relationship between chemical structure and biological activity (Cherkasov et al., 2014). For example, Rath and collaborators (Rath et al., 2022) developed 2D and 3D QSAR models to analyze how variations in hydroxyl group position and number influence polyphenol binding affinity to target proteins, aiding in the identification of potent bioactive compounds. Similarly, Lucić et al (Lucić et al., 2008) applied QSAR modeling to predict antioxidant activity across a range of polyphenols, indirectly shedding light on the structural features critical for their biological function. While QSAR models face challenges due to the structural complexity of polyphenol-protein systems, they continue to provide valuable insights into structure-activity relationships and serve as an essential tool in functional prediction and compound screening.

ML builds mathematical models that are learned from extensive datasets and iteratively improve performance, ultimately facilitating accurate predictions of interactions. A Random Forest-based RF-Score model, which was the first to use ML algorithms to predict protein-ligand binding affinity, was proposed by (Ballester and Mitchell, 2010). Random forest (RF) models and active learning variables are able to evaluate and compare the information content of different molecular representations, including protein-ligand interaction profiles (IFPs) and structure maps based on compound structure, to accurately predict the binding patterns of kinase and inhibitor (Rodríguez-Pérez et al., 2020). Aside from the Random Forest algorithm, other methods are widely utilized for protein-compound prediction. EDock-ML harnesses ML algorithms like K-Nearest Neighbor (KNN), Logistic Regression (LR), Random Forest, and Support Vector Machine (SVM) to support integrated docking to predict whether a compound is active against a target biomolecular (Chandak and Wong, 2021). Meng and Xia (Meng and Xia, 2021) introduced a few years ago PerSpect ML, a ML model based on persistent spectra. It generates a series of spectral models at different scales through a filtering process, treats the PerSpect attribute values as individual features, and accurately predicts protein-ligand binding affinity through SVM, RF, and Gradient Boosting Trees (GBT).

Although the above methods have been widely used in the study of PhPIs with some success, many challenges remain. First of all, traditional characterization methods, especially spectroscopic techniques, may be affected by sample complexity. This can result in signal superposition and background noise interference, thereby reducing data resolution and accuracy (Poklar Ulrih, 2017). When the thermodynamic method is used to determine the thermodynamic parameters of the interaction, it is affected by the experimental conditions and data processing, such as solvent selection, temperature control and model assumptions, which may lead to the uncertainty of the results (Haslam, 1996). Microscopic methods allow for the visual observation of sample morphology and structure. However, the analysis of macromolecular complexes is hindered by resolution and sampling limitations, making it difficult to capture microstructural details (Zhang, Q. et al., 2021). Spectroscopy-based NMR spectroscopy light scattering, MS, and CS also face challenges, such as NMR spectroscopy analysis of PhPIs requires longer acquisition times. Additionally, the analysis of complex samples presents difficulties due to signal overlapping and interpretation challenges (Liu et al., 2017). Similarly, light scattering techniques are impacted by sample complexity, potentially leading to signal overlaps and increased background noise, which can diminish data quality and resolution. MS and CS methods may be limited by resolution and selectivity in the identification of polyphenol-protein complexes, especially in complex food substrates where analysis can be difficult (Le Bourvellec, C. and Renard, C.M.G.C., 2012). Moreover, the simulation of large complexes using molecular docking is computationally inefficient, particularly when considering the structural diversity of polyphenols and proteins (Braun et al., 2019). MD simulations demand substantial computational resources and running time and are sensitive to initial conditions and parameters of the simulated system, which can lead to uncertainty in the results. Issues of computational efficiency and generalizability also arise with traditional ML techniques (Huang and Zou, 2010).

QSAR models offer a rapid means to predict binding affinities between polyphenols and proteins based on molecular structure, yet they too face significant challenges. Their predictive performance is highly contingent upon the quality and diversity of training datasets; insufficient or biased data frequently result in overfitting and limited generalizability (Cherkasov et al., 2014). Furthermore, conventional QSAR molecular descriptors often fail to fully capture the intricate nonlinear and synergistic interactions inherent in PhPIs, thereby constraining accuracy (Hou et al., 2020). QSAR models also typically lack interpretability and struggle to represent dynamic binding processes, limiting their utility in mechanistic studies (Baskin et al., 2016). The intrinsic structural complexity of polyphenol–protein systems, characterized by diverse binding sites and conformational heterogeneity, further complicates QSAR modeling (Tropsha, 2010).

In summary, traditional characterization techniques—such as spectroscopy, mass spectrometry, chromatography, and molecular docking—alongside MD simulations, are constrained by limited data resolution, high computational cost, and challenges in accuracy and reliability when investigating PhPIs. QSAR models offer a computationally efficient approach for predicting binding affinities based on molecular structure, yet they remain dependent on the quality and diversity of training data and often fall short in capturing complex nonlinear and dynamic interactions, as well as providing mechanistic interpretability. These limitations underscore the pressing need for methodological advancements and integrative approaches to deepen our mechanistic understanding of PhPIs.

**Table 1. PhPIs parameters: Measurement methods and DL prediction feasibility. The ability of DL to predict parameters of PhPIs is classified as feasible (√), partially feasible (√̸), or currently difficult to process (x).**

| Argument | Measurement method | Specific index | DL predictable | Instructions | Ref. |
|---|---|---|---|---|---|
| Association constant | ITC<br>UV-Vis | Ka | √ | Sequence - or structure-based regression models (e.g. GNN, Transformer) with input polyphenol/protein sequences or binding site characteristics predict Ka. | (Tsubaki et al., 2019) |
| Affinity | SPR<br>Fluorescence spectrum | $K_d$ | √ | DL models such as DeepAffinity are trained with experimental data to predict the binding free energy ΔG (directly related to Kd). | (Karimi et al., 2019) |
| Conformational change | CD<br>FTIR<br>NMR | Secondary structure change<br>Functional group change<br>Molecular vibration information<br>Binding site<br>Molecular conformation | √ | 3D-CNN analyzes spectral data to predict conformational changes<br>The time series model (LSTM) processes the MD simulation trajectory, but it needs high resolution dynamic data support. | (Abbasi et al., 2020) |
| Thermodynamic parameter | ITC<br>DSC | Enthalpy change (ΔH)<br>Entropy change (ΔS)<br>Free energy change (ΔG) | √ | DeepBindGCN fuses protein-ligand molecular characteristics with physicochemical information through GCN to accurately predict ΔG. ΔH/ΔS requires experimental calibration, and DL can assist thermodynamic trend analysis. | (Zhang et al., 2023) |
| Dynamic parameter | SPR<br>CE | Binding rate constant ($k_{on}$)<br>Dissociation rate constant ($k_{off}$) | ✓̶ | The accuracy depends on the simulated data. | -- |
| Binding site | NMR<br>Molecular docking<br>MD | Location of binding site<br>Interaction mode | √ | DeepCSeqSite only uses sequence and evolutionary features to predict protein-ligand binding sites. Attention mechanism analysis of key residues. | (Cui et al., 2019) |
| Molecular weight determination | MALDI-TOF MS<br>ESI-MS | Molecular weight<br>Binding quality | × | DL can not be directly measured molecular weight, it needs to rely on mass spectrum data; However, it can assist mass peak resolution (such as Denovo-GCN sequencing model). | (Wu et al., 2023) |
| Particle size distribution | DLS<br>SLS | Particle size<br>Aggregation state | × | The multi-scale aggregation process depends on the physical model, and DL lacks interpretability. GANs can simulate clustered images, but they are difficult to quantify. | -- |
| Surface topography | AFM<br>SEM<br>TEM | Surface morphology<br>Microstructure | × | DL can assist image segmentation (such as U-Net) to extract surface topography features, but it cannot predict the topography itself. | (Zunair and Ben Hamza, 2021) |
| Solution stability | Nephelometry<br>SEC<br>GPC | Solution stability<br>Aggregation degree<br>Molecular weight distribution | × | Stability prediction requires multiple parameters (pH, temperature, etc.), DL can be combined with environmental conditions to classify stability levels, but it is difficult to replace experiments. | -- |
| Interfacial | Rheology of interface | Interfacial viscoelasticity | × | Interfacial rheology depends on multi- | -- |

| | | | |
|---|---|---|---|
| viscoelasticity | dilation | film strength | physical field coupling (fluid mechanics + molecular interaction), DL is difficult to model. The hybrid model is being explored |

### 3.2. The application of DL in elucidating PhPIs

In recent years, the rapid advancement of AI, particularly the emergence of DL algorithms, has provided new insights to solve challenges faced by traditional characterization methods. DL aims to learn from existing interaction data and its powerful computational capabilities to identify complex features and patterns, thereby revealing the mechanism behind PhPIs (Hong et al., 2024; Moon et al., 2024a; Wang, D.D. et al., 2024). For instance, Watanabe and coworkers (Watanabe et al., 2021) introduced a DL approach that integrates protein and compound features, as well as various types of interaction set data to accurately predict protein-compound interactions.

DL techniques can be employed to construct highly adaptable models for processing complex spectral data. Through the feature extraction and learning capabilities inherent in deep neural networks, signal and noise can be more effectively distinguished, enhancing data resolution and accuracy (Liu, X. et al., 2024). Furthermore, algorithms such as convolutional neural networks (CNNs) perform well in processing image and sequence data and can be used to overcome the resolution constraints of microscopy, capturing microscopic structural details. This automated data processing method not only saves researchers time and effort but also mitigates the impact of human factors on outcomes and improves the reliability and reproducibility of studies (Liu, Z. et al., 2021).

Many traditional studies on PhPIs rely on empirical methods, which yield surface-level insights that hinder deep exploration of the underlying mechanisms and principles. In contrast, DL technology employs the power of big data mining and knowledge acquisition to uncover potential patterns and laws. The application of DL technology to the study of PhPIs has the potential to reveal novel mechanisms of action between these two classes of molecules, as well as to provide a fresh perspective on interactions at a profound level. This algorithm-driven approach to mechanism interpretation and relationship discovery not only informs experimental design and result interpretation but also propels research forward. Moreover, when combined with bioinformatics and computational biology methodologies, DL can explore the molecular-level interaction mechanisms of PhPIs. For example, DL-based models can simulate spatial structures and dynamic changes between polyphenols and proteins, thus elucidating the PhPIs mechanisms (Moen et al., 2019).

While the initial training of DL models may require a significant investment of time and computational resources, the subsequent inference and prediction processes are typically highly efficient and fast. In particular, the advances in hardware technology and the optimization of DL algorithms have contributed to this improvement. Consequently, the computational costs associated with DL are gradually diminishing, rendering it increasingly feasible for practical applications in the study of PhPIs (Zhang et al., 2018).

## 4. Applications scenarios of DL in the analysis of PhPIs

As previously stated, extensive investigation has been conducted on PhPIs. The rapid growth of AI and the protein structural data has facilitated broader prospects for protein-related research assisted by DL algorithms. The application of DL algorithms is regarded as a valuable approach for elucidating the interaction pattern and impact of PhPIs. Further insights into the underlying properties of PhPIs, including chemical attributes, spatial conformations, bioavailability, and binding sites of their complexes, which are currently a research focus in this field, can also be provided.

The process of studying PhPIs with the aid of DL algorithms generally follows a structured process (Crampon et al., 2022). An overview of this workflow, including data acquisition, preprocessing, model construction and training, as well as performance evaluation, is illustrated in **Figure 4**. Initially, data on proteins, polyphenols, and their interactions are gathered from various databases. This information is then processed to transform it into computer-readable data representations. Through subsequent feature extraction and selection steps, key information relevant to downstream analysis is identified. DL algorithms, which utilize either protein or polyphenol sequences or structures as inputs, are used to establish the relationship function of interactions. Following an extensive training phase aimed at achieving high prediction accuracy, the model is debugged, tuned, and performance evaluated. Finally, the refined model is deployed to predict PhPIs involving unknown proteins and/or polyphenols.

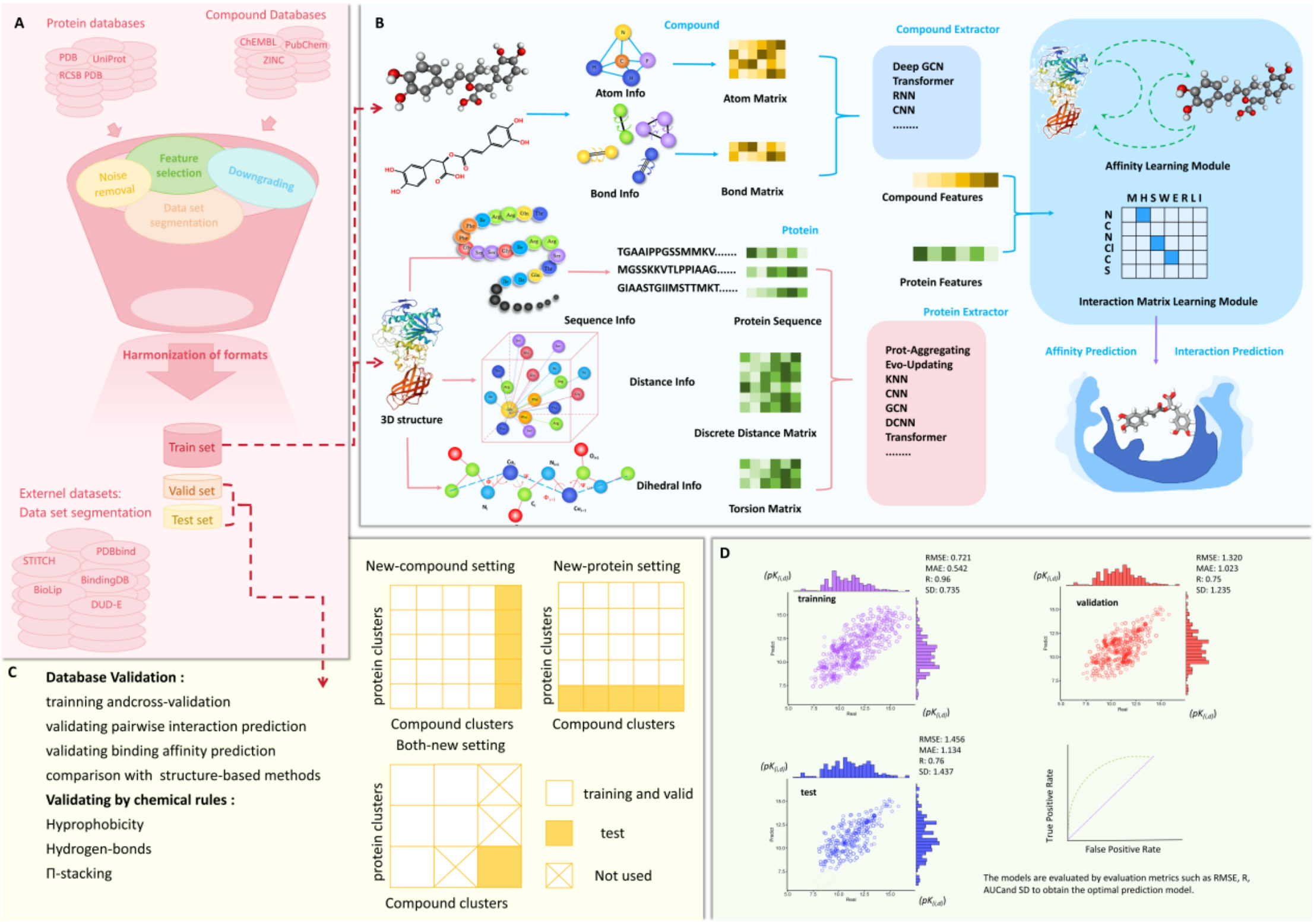

**Figure 4: Workflow and methodology overview for protein-compound interaction prediction in PhPIs:** (**A**) Data preprocessing involves the integration and normalization of protein and compound datasets from sources such as PDB, UniProt, ChEMBL, and ZINC. Noise removal, feature selection, and segmentation ensure compatibility with downstream ML tasks. External datasets are utilized for independent validation and testing. (**B**) The predictive model combines graph-based algorithms and neural network models (e.g., Deep GCN, CNN, and Transformer) to extract features from protein sequences and structures as well as compound atom and bond matrices. Interaction prediction is achieved through affinity learning and interaction matrix learning modules. (**C**) Validation strategies include cross-validation, affinity prediction, and chemical rule-based assessments (e.g., hydrophobicity and hydrogen bonding), with comparisons to structure-based methods enhancing reliability. (**D**) Model performance is evaluated using metrics such as RMSE, $R^2$, and AUC-ROC across training, validation, and test sets. Distribution plots highlight the accuracy of binding affinity (Kd) predictions.

## 4.1. Data acquisition and pre-processing

Typically, approximately 80% of the time dedicated to constructing AI-based predictive models is allocated to data preparation and processing (Vamathevan et al., 2019). Abundant data enables the data model to gain a more comprehensive understanding of the characteristics of proteins and polyphenols. **Table 2** provides a summary of pertinent data related to proteins, enzymes, small molecule ligands, compound-protein interaction (CPI), protein-ligand binding affinities (PLBA) and other relevant data. Notably, the Protein Data Bank (PDB) is the most comprehensive three-dimensional protein structure database in the world, which was established in 1971. The PDBbind database (Wang et al., 2005) is comprised of a collection of all protein-ligand complexes archived in the PDB, along with experimentally measured binding affinity data. Information on the chemical structures of numerous organic compounds, as well as biochemical experimental data, is contained within PubChem. These databases are regarded as invaluable resources for researchers involved in construction and training of models for PhPIs. Additionally, they can also be employed as training and test sets to evaluate model performance. To train their model, Kroll et al. (Kroll et al., 2023) used the GO annotation database of UniProt IDs to create a dataset with experimentally confirmed enzyme-substrate pairs. A total of 18351 enzyme-substrate pairs with experimental evidence of binding were extracted, consisting of 12156 unique enzymes and 1379 unique metabolites. A total of 274,030 enzyme-substrate pairs with evidence of phylogenetic inference were also extracted, indicating that these enzymes are evolutionarily closely related to enzymes associated with the same reaction. Moreover, numerous researchers have aggregated the data filtered or predicted by their models into new databases for further utilization. One such is PrePCI database, which was developed by (Trudeau et al., 2023).

For structural analysis, PDB offers 3D templates of existing polyphenol-protein complexes (Berman et al., 2000). SwissModel enables researchers to build protein models when structural data is unavailable (Waterhouse et al., 2018). Experimental data on polyphenol-protein binding is available through BindingDB (Liu et al., 2006) and ChEMBL (Gaulton et al., 2016). The ZINC database (Sterling and Irwin, 2015) facilitates the screening of polyphenol compounds for molecular docking studies, and DALI (Holm, 2022) compares protein 3D structures to identify similar regions, which may indicate shared binding sites in related proteins. Docking tools, including AutoDock (Morris et al., 2009) and SwissDock (Grosdidier et al., 2011), predict binding affinities, while GROMACS (Abraham et al., 2015) and AMBER (Salomon-Ferrer et al., 2013) simulate MD. Visualization tools like PyMOL (Seeliger and de Groot, 2010) and Chimera (Pettersen et al., 2004) allow researchers to analyze binding sites more effectively.

Additional resources further enrich PhPIs research. PubChem provides bioactivity data on polyphenols and their targets (Kim et al., 2015). InterPro (Blum et al., 2020) supports the analysis of protein families and domains, while SwissTargetPrediction (Daina et al., 2019) forecasts potential protein targets of polyphenols using multidimensional data. By combining these tools with experimental data from sources such as BindingDB and ChEMBL, researchers can build a more comprehensive framework to explore PhPIs.

**Table 2: Protein and compounds information databases, along with active URL.** A "browsable" website refers to a website on which the user can explore the contents of the database without downloading files or using an Application Programming Interface (API). A database is denoted with "download available" if the entirety of the database is available to download at once.

| Database | Description | Browsable Website Available | Download Available | API Available | URL |
|---|---|---|---|---|---|
| Binding MOAD | A database providing information on protein-ligand crystal structures with detailed annotations. | + | + | - | http://www.BindingMOAD.org |
| BindingDB | A database specializing in molecular binding data. | + | + | + | https://www.bindingdb.org/rwd/bind/index.jsp |
| BioLiP | A semi-managed database of biologically relevant ligand-protein binding interactions. | + | + | + | https://seq2fun.dcmb.med.umich.edu/BioLiP/index.cgi |
| BRENDA | A database providing detailed information on enzyme function and properties. | + | + | + | https://www.brenda-enzymes.org/ |
| CASF-2016 | A database specifically designed for the evaluation of docking and scoring methods for PLBA. | - | + | - | http://www.pdbbind-cn.org/casf.asp/ |
| ChEMBL | a manually curated database of bioactive molecules with drug-like properties. | + | + | + | https://www.ebi.ac.uk/chembl/ |
| DUD-E | A database for evaluating molecular docking and algorithmic modeling. | + | + | - | https://dude.docking.org/ |
| DrugBank | a database for drugs, drug actions and drug targets. | + | + | + | https://go.drugbank.com/ |
| eF-site | a database of protein functional sites. | + | - | - | https://pdbj.org/eF-site/ |
| eModel-BDB | a database of comparative structure models of drug-target interactions from the Binding Database. | + | + | - | https://brylinski.org/emodel-bdb-0 |
| FireDB | A database of proteins with structures, ligands and annotated functional site residues. | + | + | + | https://firedb.bioinfo.cnio.es/Php/FireDB.php |
| LigBase | a database of ligand binding proteins aligned to structural templates. | + | - | - | https://modbase.compbio.ucsf.edu/ligbase/ |
| Phenol-Explorer | A database containing polyphenol content in foods, *in vivo* metabolism and pharmacokinetics | + | + | - | http://www.phenol-explorer.eu |
| PDB | A database providing 3D structural data of biological molecules. | + | + | - | https://www.wwpdb.org/ |
| PDBbind | A database focused on curating and sharing binding affinity information for protein-ligand interactions. | + | + | - | https://www.pdbbind-plus.org.cn/ |
| PDBSite | a database of protein functional site. | + | - | - | http://wwwmgs.bionet.nsc.ru/mgs/gnw/pdbsite/ |
| PiSite | A PDB-based database for searching for interaction sites in protein sequences. | + | + | - | https://pisite.sb.ecei.tohoku.ac.jp |
| PrePCI | A structure- and chemical similarity-informed database of predicted protein compound interactions. | + | - | - | https://honiglab.c2b2.columbia.edu/prepci.html |
| PubChem | An open chemistry database mostly contains small molecules (also large molecules) with their properties and active. | + | + | + | https://pubchem.ncbi.nlm.nih.gov/ |
| sc-PDB | An annotated database of druggable binding sites from the Protein Databank | + | + | - | http://bioinfo-pharma.u-strasbg.fr/scPDB/ |
| STITCH | A database for searching for known or predicted interactions between compounds and proteins | + | + | + | http://stitch.embl.de/ |
| UniProt | A database providing comprehensive information on protein sequences and functions across various species. | + | + | + | https://www.uniprot.org/ |
| ZINC | A database of small molecule structures that can be used for molecular docking | + | - | + | https://zinc.docking.org/ |

Data with highly accurate, representative, and uniformly formatted information is essential for enhancing the robustness and efficacy of the model, facilitating the precise prediction of PhPIs (Mohammed et al., 2025). In short, both the quantity and quality of data are considered fundamental for improving model generalization performance (Yang et al., 2018). Given that computer systems can only interpret numeric vectors, raw data must undergo transformation into formats recognizable by computers after acquisition (Yang, K.K. et al., 2019), followed by preprocessing and feature selection (**Fig. 5A**). To ensure data reliability, numerous researchers employ various means to enhance both the quantity and quality of data. For instance, (Yang, Y. et al., 2020) gathered multiple datasets of different sequences for the purpose of assessing model robustness, while (Li, H. et al., 2022) developed a DL-based data preprocessing model, AutoClass, which filters out various noises and artifacts by integrating two deep neural network components-an autoencoder and a classifier. This integration enables consistent and significant improvement of data quality and downstream analysis, including differential representation and clustering.

Feature extraction during preprocessing represents a critical data processing step that profoundly influences the construction of downstream predictive models, serving as a key factor in controlling data quality. The process of screening out significant features serves to reduce the dimensionality of raw data, thereby accelerating the speed of training. Furthermore, it minimizes interference noise, mitigates overfitting risks and enhances the efficacy of the model. MotifCNN and MotifDCNN are feature extraction methods constructed based on CNN, which are used to detect structural motifs related to protein folding, extract folding-specific features, and effectively improve the discrimination ability of specific folding features (Li and Liu, 2019).

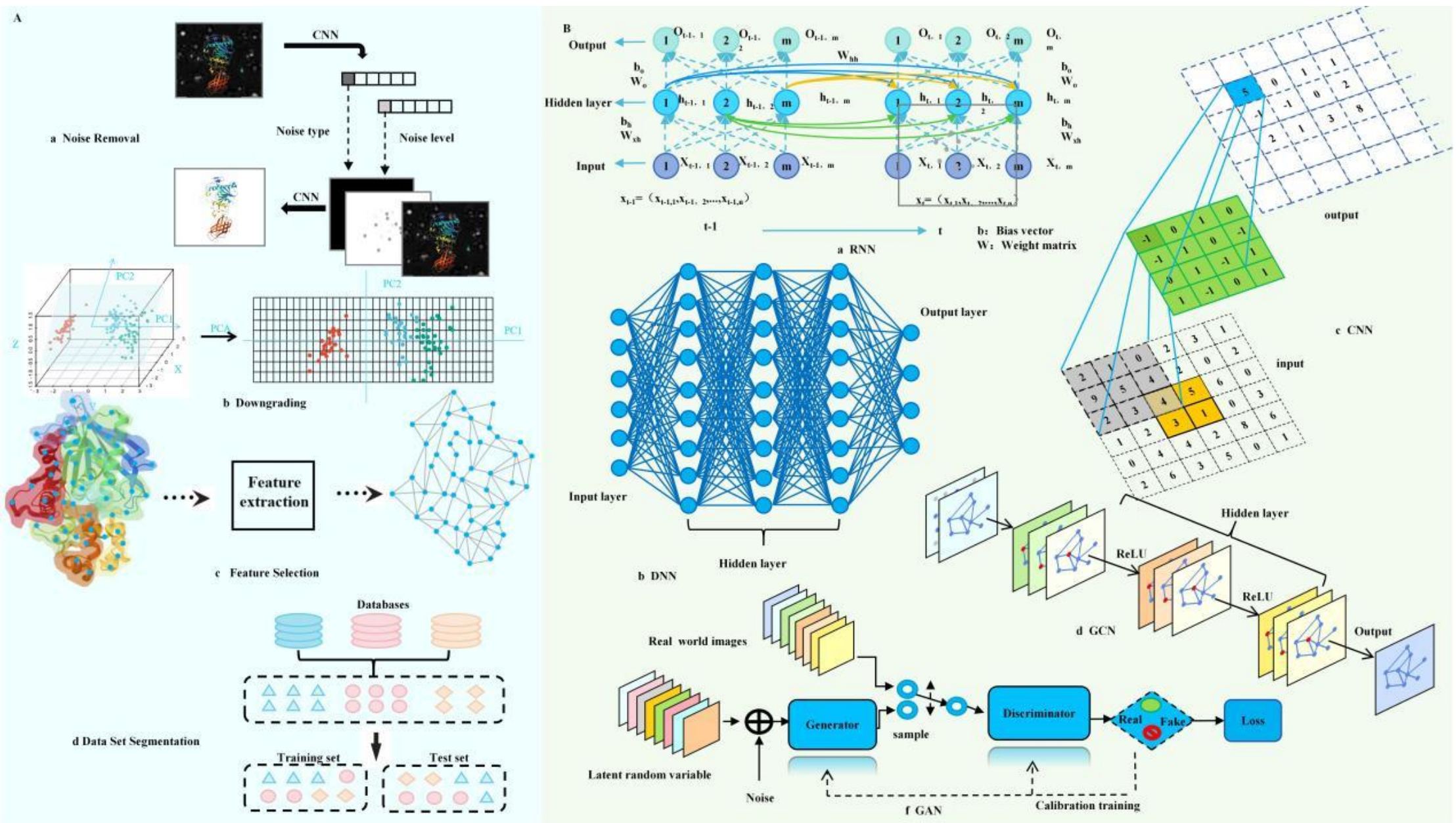

**Figure 5: Neural network architecture and processing workflow overview.** (A) Demonstrate workflows for preprocessing, feature extraction, and data set segmentation in a computational framework. (a) demonstrates noise removal using a convolutional neural network (CNN), which identifies noise types and levels for correction. (b) shows the dimensionality reduction of principal component analysis (PCA) and its mapping to the feature

space. (c) depicts feature selection and network topology construction based on extracted features. (d) depicts the partitioning of the data set into training subsets and test subsets for model evaluation. **(B)** Schematic of different neural network architectures. (a) illustrates a recurrent neural network (RNN) that emphasizes cyclic connections, hidden states, and weight updates. (b) represents a deep neural network (DNN) with fully connected layers for hierarchical feature learning. (c) highlights the convolutional neural network (CNN) structure, showing the process of layered feature extraction through convolutional layers and pooling layers. (d) Visualizes generative adversarial networks (Gans), consisting of generators and discriminators that optimize the output through calibration training. (e) introduces a graph convolutional network (GCN), emphasizing the processing of graph structured data through convolutional layers.

### 4.2. Model building and training

The goal of model construction is to autonomously identify data features from existing datasets, knowledge, or experience and to employ algorithmic training to uncover potential relationships and patterns. Through this process, a model with precise predictive and decision-making capabilities for unknown data can be developed. The algorithm utilized by the model must select the optimal solution based on specific task requirements (LeCun et al., 2015). For example, CNNs are well-suited for image recognition analysis, whereas classification and regression tasks often employ multi-layer perceptrons (MLPs). For natural language processing (NLP) endeavors, recurrent neural networks (RNNs) or TRANSFORM can be selected. The pre-selected model algorithm is continuously trained on a large training set to adjust the model parameters to obtain better fitting ability and improve performance. Throughout the training phase, the model's performance on both known and unknown data can be assessed to gauge its fitting and generalization capabilities. A robust model should exhibit strong fitting on training data and effective generalization to unknown data. Therefore, during model training, it is essential to strike a balance between the fitting and generalization capabilities to ensure the effectiveness and reliability of the model in practical applications (Dietterich, 2000). Achieving this balance involves adjusting model complexity, regularization parameters, and dataset size and quality. To enhance model accuracy and stability, multiple models are sometimes constructed simultaneously and then integrated to assure that the model has an overall performance (Srivastava et al., 2014).

A DL algorithm is a neural network-based ML algorithm that simulates the neural network structure inspired by the human brain to learn features and make decisions across multiple levels of neurons and weighted connections (Zheng et al., 2020). Such algorithms include Graph Neural Network (GNN), RNN, CNN, Graph Convolutional Neural Network (GCN), deep neural network (DNN) and Generating Adversarial Neural Network (GAN) (**Fig. 5B**). In recent years, major breakthroughs have been made in the fields of image recognition, natural language processing, and speech recognition. Due to its significantly enhanced ability to explore potentially complex information in data features, effective models based on DL algorithms are proposed to predict potential protein-compound interactions, as shown in **Table 3.** Wan et al (Wan et al., 2019) introduced a DNN classifier that utilizes multimodal DNNs. Compound-protein interaction probabilities are predicted by DeepCPI through unsupervised learning of feature embeddings, which represent compounds and proteins in low-dimensional feature spaces. This approach employs semantic analysis and the Word2vec method on a comprehensive database of compounds and proteins. While many of these models rely on high-quality protein data, some efforts have been made to address their limitations. Structure-free models, which are based on overall similarity and graphical representations of compound molecular features, have been proposed as potential alternatives.

For instance, CPI-IGAE transforms protein heterogeneity graphs into chi-square graphs with directed and weighted edges, learning low-dimensional representations of compounds and proteins in an end-to-end fashion from these chi-square graphs (Wan et al., 2022). Similarly, Wang et al (Wang, Z. et al., 2024) extracted amino acid sequence features from polyphenol molecular maps and proteins using two specific convolutional networks, encoding them into computer-recognizable feature matrices before constructing a BANPPI model for interaction prediction via a bilinear attention network that elucidates the relationships between danphenol and various proteins such as bovine serum albumin, bovine β-lactoglobulin, egg ovalbumin, zein alcohol-soluble protein, bovine α-lactalbumin, lactoferrin among others; this can inform the design of polyphenol-protein complex delivery systems. Additionally, the MONN model combines supervised learning methods, where a graph convolution module and a CNN module are integrated to extract atomic and residue features from input molecular maps and protein sequences respectively. Pairwise interaction matrices are defined to describe non-covalent interactions between input compounds and protein residues. For a non-hydrogen atom compound containing $Na$ atoms and a protein with Nr residues, the interaction matrix $P$ is defined as a $Na \times Nr$ binary matrix. Where each element $P_{ij}$ ($i = 1, 2, 2, ...; Na$ and $j = 1, 2, ...; Nr$) indicates whether there is a non-covalent interaction between the $i$ th atom of the compound and the $j$ th residue of the protein (1 is present, otherwise 0). Interaction sites can be inferred from this matrix by maximizing the rows or columns. And it outperforms other ML methods in predicting binding affinity (Li et al., 2020).

**Table 3: DL models and their constructing modes and code availability, and how proteins and compounds are imported.**

| Method Name | Constructing modes and code availability | Input protein features | Input compound features | Binding affinity regression | Binding sites | Ref. |
|---|---|---|---|---|---|---|
| DeepAffinity | attention RNN-CNN model for semi-supervised learning of proteins and compounds<br>Code: https://github.com/Shen-Lab/DeepAffinity | Sequences | SMILES strings | + | - | (Karimi et al., 2019) |
| DeepCPI | DeepCPI: latent semantic analysis for compounds and Word2Vec for proteins<br>Code: https://github.com/FangpingWan/DeepCPI | Sequences | Sequences | - | - | (Wan et al., 2019) |
| CPI_prediction | GNN for compounds and CNN for proteins and then joint a neural attention mechanism<br>Code: https://github.com/masashitsubaki/CPI_prediction | Sequences | 2D graphs | - | + | (Tsubaki et al., 2019) |
| MONN | MONN: GCN for compounds and CNN for proteins<br>Code: https://github.com/lishuya17/MONN | Sequences | 2D graphs | + | + | (Li et al., 2020) |
| DrugVQA | DrugVQA: a dynamic CNN incorporating with a sequential attention module for proteins and a bidirectional LSTM incorporating with multi-head attention for compounds<br>Code: https://github.com/prokia/drugVQA | 2D distance map | SMILES strings | - | + | (Zheng et al., 2020) |
| TransformerCPI | TransformerCPI: GCN for compounds and Word2Vec for proteins and then use multi-headed attention layers in the transformer decoder<br>Code: https://github.com/lifanchen-simm/transformerCPI | Sequences | SMILES strings | - | + | (Chen et al., 2020) |
| DeepCDA | DeepCDA: a combination of CNN and LSTM for proteins | Sequences | SMILES | + | + | (Abbasi et al., 2020) |
| | and compounds and then a two-side attention layer, respectively<br>Code: https://github.com/LBBSoft/DeepCDA | | strings | | | |
| MDeePred | MDeePred: CNN for proteins and DNN for compounds<br>Code: https://github.com/cansyl/MDeePred | Multiple 2D vectors | Fingerprints | + | - | (Rifaioglu et al., 2021) |
| GraphDTA | GraphDTA: GCN, GAT, GIN and a combined GAT-GCN architecture for compounds and CNN for proteins<br>Code: https://github.com/thinng/GraphDTA | Sequences | 2D graphs | + | - | (Nguyen, T. et al., 2021) |
| MultiDTI | MultiDTI: two CNN blocks for proteins and compounds and added a heterogeneous network of compounds, proteins, side effects and diseases<br>Code: https://github.com/Deshan-Zhou/MultiDTI | Sequences | SMILES strings | - | - | (Zhou et al., 2021) |
| DeepConv-DTI | MolTrans: Transformer for compounds and proteins<br>Code: https://github.com/GIST-CSBL/DeepConv-DTI | Sequences | SMILES strings | - | - | (Huang et al., 2021) |
| MATT_DTI | MATT_DTI: 1-D CNN for compounds and proteins<br>Code: https://github.com/ZengYuni/MATT_DTI/ | Sequences | SMILES strings | + | - | (Zeng et al., 2021) |
| DeepDTAF | DeepDTAF: combined the dilated convolution with traditional convolution for compounds and proteins<br>Code: https://github.com/KailiWang1/DeepDTAF | 1D Sequences | SMILES strings | + | - | (Wang et al., 2021) |
| Multi-PLI | Multi-PLI: CNN for proteins and compounds<br>Code: https://github.com/Siat-Code/Multi-PLI/ | Sequences | SMILES | + | - | (Hu et al., 2021) |
| Point-Cloud | PointNet/PointTransformer ： compounds and proteins transformer to point cloud<br>Code: https://github.com/wyji001/Point-Cloud | the point cloud | the point cloud | + | - | (Wang, Y. et al., 2022) |
| AttentionSiteDTI | AttentionSiteDTI: GCNN for after simulation-based model preprocessed protein and ligand | 3D structures | SMILES strings | - | + | (Yazdani-Jahromi et al., 2022) |
| | Code: https://github.com/yazdanimehdi/AttentionSiteDTI | | | | | |
| ELECTRA-DTA | ELECTRA-DTA: CNN for proteins and compounds<br>Code: https://github.com/IILab-Resource/ELECTRA-DTA | the amino acid sequences of proteins | SMILES strings | + | - | (Wang, J. et al., 2022) |
| KDBNet | KDBNet: GNN for proteins and compounds<br>Code: https://github.com/luoyunan/KDBNet | 3D structures | 3D structures | + | - | (Luo et al., 2023) |
| FeatNN | FeatNN: Prot-Aggregation and Evo-Updating for protein and GCN for compounds<br>Code: https://github.com/StarLight1212/FeatNN | Sequences and 3D structures | Molecular graph | + | - | (Guo et al., 2023) |
| CmhAttCPI | CmhAttCPI: GCW for compounds and CNN for proteins<br>Code: https://github.com/wangmeng-code/CmhAttCPI | Sequences | Graph representation | - | + | (Wang, M. et al., 2024) |
| BCMMDA | BCMMDA: CNN for feature learning<br>Code: https://github.com/Huangzhuo119/BCMMDA | Sequences | SMILES | - | - | (Huang, Z. et al., 2024) |

### 4.3. Performance evaluation of PhPIs prediction models

The objective of model evaluation is to select models with strong generalization ability to fulfill prediction tasks. Model evaluation methods include Hold-out, Cross Validation, Bootstrap, and others, which comprehensively evaluate model quality and guide model tuning and refinement to enhance its performance and generalization ability (Kohavi, 1995). Among them, the most commonly used method is Cross Validation (Nguyen, T.B. et al., 2021), which splits the dataset into $k$ subsets, one of which is held as the validation set, while the remaining $k-1$ subsets are used for training. This process is repeated $k$ times, each time with a different subset as the validation set and the rest for training. Finally, the results of each validation are averaged to obtain the final model performance evaluation. Assessing the accuracy and precision of the model requires the establishment of appropriate performance indicators. Performance indicators such as accuracy, precision, recall, F1 score, enrichment factor (EF), and area under the curve (AUC) of the ROC curve are used to evaluate model performance in classification tasks, particularly for binary classification problems. Mastropietro et al (Mastropietro et al., 2023) recently assessed the predictive performance of six GNN-based models using root-mean-square error (RMSE), while D. Zhou et al. (2022) employed a trio of metrics-area under the ROC curve, precision-recall curve (PRC), and enrichment factor (EF)-to evaluate the prediction performance of Deffini network model (Zhou et al., 2022). On the Kernie dataset, three clustering cross-validations were conducted. The evaluation results showed that the average AUC_ROC was 0.985, AUC_PRC was 0.857, and EF values were 44.8% at 1%, 17.2% at 5%, and 9.1% at 10%, respectively, which were superior to those of other DL methods.

## 5. Major Gaps in DL-Driven PhPIs Research

### 5.1. Limitations on data quality and quantity

While these experimental and computational methods provide powerful tools for studying PhPIs, several challenges remain. One major challenge is the diversity of polyphenols and proteins, which makes it difficult to develop universal models for all interactions (Ozdal et al., 2013). Additionally, the dynamic and transient nature of PhPIs, influenced by factors such as pH, temperature, and ionic strength, further complicates prediction accuracy (Euston, 2021). Thus, the efficacy of data-driven models is greatly dependent on the quantity and quality of the training data.

The quantity of available data directly influences the training and performance of DL models. Predictive models are often unable to make accurate predictions on data not covered by their training datasets. Despite the substantial volume of data on protein-ligand interactions accumulated in the bioinformatics field, it remains relatively limited. Since features must be acquired and potential relationships from sufficient data samples, the generalization ability of the algorithm is constrained by a lack of sufficient samples (Rifaioglu et al., 2018). This limitation can result in overfitting or underfitting of the model, thereby reducing its accuracy and reliability. Henryl *et al.* trained ProteinMPNN using varying sizes of training sets, and their findings indicated that models trained on larger datasets exhibited superior accuracy and ROC-AUC scores. However, merely augmenting the data volume may not enhance model performance, as data diversity is of greater significance than sheer quantity in furnishing comprehensive feature information.

Data quality also poses a significant challenge. It serves as the foundation for ensuring that models effectively learn the correct and crucial features, which directly impacts the accuracy, integrity and reliability of the model (Fan and Shi, 2022). PhPIs data typically stems from experimental measurements or computer simulations. These datasets are susceptible to a variety of factors, including experimental errors, noise, labeling errors, and incomplete data. The data collected in public databases are often generated by different biological analyses, methods, or conditions, which makes direct comparison between them challenging (Özçelik et al., 2023). Moreover, contradictions may be present among multiple datasets on the same topic. Incomplete data, defined as data points that are absent or data sets that exhibit inter-sample bias, can impair a model's generalization ability and stability. Quality issues like mislabeling or inconsistency further diminish the reliability of DL models. Imbalances in data sampling, whereby certain protein or ligand classes are oversampled during feature extraction and selection, can lead to poor predictions for these classes during model training and inference, thereby compromising the model's comprehensiveness and accuracy (Yang, J. et al., 2020). Consequently, acquiring accurate and comprehensive valid data remains an ongoing challenge.

### 5.2. Learning from Protein Dynamics

Protein dynamics are essential for understanding the complex relationship between structure and function. However, studying these dynamic processes in depth remains challenging (Henzler-Wildman and Kern, 2007). Traditional methods like X-ray crystallography provide high-resolution images of protein structures but cannot capture the conformational flexibility needed during ligand binding (Torrens-Fontanals et al., 2020). Advanced techniques such as nuclear magnetic resonance spectroscopy (NMR) and cryo-electron microscopy (Cryo-EM) partially address this issue. However, NMR is limited by the size of the molecules, while Cryo-EM lacks the temporal resolution and submillisecond-level dynamic capture capabilities needed (Voss et al., 2021). Research has shown that analyzing protein assemblies, rather than single static structures (Ruscio et al., 2009), can predict thermal stability (Peccati et al., 2023), identify remote conformational effects from substrate redirection (Acevedo-Rocha et al., 2021), and highlight regions that promote dynamic reactivity (Bonk et al., 2019). As a result, there has been a push to develop tools that generate dynamic information from sequence or structural data without requiring heavy computation (Vander Meersche et al., 2021). This need is driving the development of new methods based on dynamic data (Zheng, L.-E. et al., 2023).

MD simulations offer new opportunities for studying protein dynamics. Tools like GROMACS enable MD simulations to track protein motion at atomic resolution. These simulations provide detailed insights into protein-ligand interactions by calculating atomic positions and interactions over time (Abraham et al., 2015). However, the accuracy of MD simulations depends heavily on force field parameters and computational resources. Large or highly dynamic systems require significant computational time (Shaw et al., 2010). To overcome these challenges, researchers have proposed hybrid methods combining coarse-grained models or enhanced sampling techniques to improve simulation efficiency (Jung et al., 2024). Currently, most protein dynamics prediction methods still rely mainly on static structure or single sequence data, and fail to make full use of dynamic information provided by MD simulation (Chen, Y. et al., 2024). The fundamental reason lies in the complex characterization of MD data: high-dimensional tracks (atomical-level time series) lead to difficulty in feature extraction (Alquier et al., 2020), and the computational cost of all-atom simulations limits large-scale data generation (Mirarchi et al., 2024). To solve this problem, a feasible strategy is to extract the key features from the MD locus, such as the partial MD simulation that can be replaced by B-factor subclass feature extraction (Song et al., 2024), or the representative conformation obtained by cluster analysis and Markov state modeling (Arbon et al., 2024). However, the generality of these methods to multiple protein datasets remains to be further validated.

Another major barrier to integrating MD data is the lack of consistent, high-quality datasets. Simulation results often come in inconsistent formats, are large in size, and are difficult to share (Abraham et al., 2019). Additionally, the sensitivity of force field parameters and settings complicates dataset construction (Serafeim et al., 2016). To address these challenges, applying the FAIR principles—Findability, Accessibility, Interoperability, and Reproducibility—can regulate data publishing and significantly improve data usability, thus advancing dynamics research (Wilkinson et al., 2016). Future research should prioritize building high-quality dynamic datasets for specific proteins or protein families (Tiemann et al., 2024). Additionally, exploring the potential for knowledge transfer between different protein families could accelerate progress (Durumeric et al., 2023).

**5.3. Missing Gold-Standard Data Sets**

The lack of gold standard data sets in PhPIs research is a core challenge that limits the use of ML and DL techniques. Unlike studies on protein-ligand or protein-protein interactions, polyphenols' diversity and complex binding behavior make it difficult to create standardized baseline data sets. Existing experimental data are often incomplete, inconsistent, or non-reproducible. This limits the ability to validate model predictions and apply universal conclusions (Huang and Zou, 2010).

The absence of gold standard data directly impacts model generalization and prediction accuracy. ML and DL technologies require large, high-quality data sets to extract meaningful features and make generalizations (Mohammed et al., 2025). If the training data are biased or insufficient, the model may perform well on a specific dataset but fail to predict accurately for new molecular structures or complex biological systems (Kretschmer et al., 2025). For instance, models may struggle to capture dynamic binding behavior in polyphenols with different molecular weights or proteins in various folded states (Butler et al., 2018). Moreover, converting complex polyphenol-protein systems into formats for ML models (e.g., SMILES or molecular graphs) may lead to the loss of critical information, especially when considering MD or nonlinear activity landscapes (Polash et al., 2019).

Additionally, the lack of gold standard data sets hinders the interpretability and validation of model results. DL models are often viewed as "black boxes," requiring reliable experimental data for validation (Ribeiro et al., 2016). However, current experimental data are limited both in quantity and quality, making verification difficult. This limitation has slowed the broader adoption of ML and DL techniques in PhPIs research (Ching et al., 2018). In contrast to widely used benchmarks in computer vision, such as MNIST or ImageNet, creating data benchmarks for protein engineering is much more challenging. It requires specialized expertise and overcoming technical barriers to experimental validation (Butler et al., 2018).

**5.4. DL for low-N data**

In low-data environments, DL faces several hurdles when interpreting PhPIs. One major challenge is the lack of sufficient training data. This often leads to overfitting, where the model learns noise instead of universal patterns (Yang, K. et al., 2019). This issue is especially pronounced in high-dimensional PhPIs data, where limited samples make it hard to capture complex relationships (van Tilborg et al., 2024). The high dimensionality of biological data also restricts the ability of traditional neural networks to extract meaningful patterns when data is scarce. Data imbalance is another issue (Angermueller et al., 2016). Certain polyphenols or protein classes often have fewer samples, which weakens the model's predictive performance in these categories (van Tilborg et al., 2024). Additionally, noisy data and inconsistent experimental results make training even more difficult. This is particularly problematic in PhPIs studies, where experimental errors can skew results (Walters and Barzilay, 2021). Low-data settings also suffer from a lack of standardized datasets and evaluation metrics, making it difficult to compare and validate methods (van Tilborg and Grisoni, 2024). Finally, the computational complexity of DL models poses a challenge. Developing lightweight models that work effectively with small datasets remains a pressing need (Brigato and Iocchi, 2021).

To overcome these challenges, researchers have introduced several strategies. Transfer learning is a key approach. Models pre-trained on large datasets, like AlphaFold or ESM, are adapted to low-data tasks, significantly improving the prediction of PhPIs (van Tilborg et al., 2024). Active learning prioritizes labeling the most informative samples. This reduces the burden of data annotation while improving model performance, especially in binding predictions (van Tilborg and Grisoni, 2024). Data augmentation techniques, such as SMILES transformations or virtual screening, expand training datasets and improve model generalization. These methods are particularly effective in addressing data scarcity in cheminformatics (van Tilborg and Grisoni, 2024). Meta-learning is another promising strategy (Yue et al., 2025). It uses prior knowledge from similar tasks to help models quickly adapt to new tasks with limited data. This makes it particularly useful for PhPIs modeling. Domain knowledge integration, such as incorporating physical or chemical property insights, improves both model performance and interpretability. Hybrid models that combine data-driven and domain-specific approaches have shown potential in addressing biological challenges (van Tilborg and Grisoni, 2024). Semi-supervised and self-supervised learning methods also help (Rives et al., 2021). By leveraging large amounts of unlabeled data, these techniques significantly enhance model performance even with limited labeled datasets.

Although these strategies address some low-data challenges, their success often depends on the specific application. Future efforts should prioritize creating standardized datasets, designing lightweight models, and developing frameworks that integrate these strategies. These steps will maximize the potential of DL in interpreting PhPIs in data-limited scenarios.

## 6. Future directions

### 6.1. Trustworthiness and Explainable AI

The increasing adoption of AI and ML models, particularly complex models such as DL and neural networks, has brought the issues of transparency and interpretability to the forefront. While DL models excel in predictive accuracy, their complexity and "black box" nature hinder researchers and users from understanding the decision-making process, which challenges the trustworthiness and practical application of these models (Alizadehsani et al., 2024). The "black box" issue has become a central challenge in the development of AI.

To address this, Explainable AI (XAI) has emerged to enhance the transparency and interpretability of AI models, particularly in the context of complex DL systems (Zhou et al., 2023). XAI employs various methods, such as Saliency Maps, SHAP (Shapley Additive Explanations), and LIME (Local Interpretable Model-agnostic Explanations), to help users understand the reasoning behind model predictions, revealing the dependencies between input data and model outcomes (Alizadehsani et al., 2024). These techniques improve model transparency and accountability. In drug discovery, for instance, the "black box" nature of AI and ML models has significantly hindered researchers' trust in model outcomes, especially in critical tasks such as drug target identification and toxicity prediction. XAI methods like SHAP and LIME allow researchers to gain a clearer understanding of how AI models analyze genomic data and identify drug targets, enhancing the credibility of model predictions and ensuring alignment with biological and chemical principles (Alizadehsani et al., 2024).

In compound design, techniques such as Generative Adversarial Networks (GANs) and Variational Autoencoders (VAEs) generate potential drug-like molecules but often lack intuitive design logic. Researchers can address this by applying attention mechanisms to highlight key functional groups and utilizing Saliency Maps for visualization, thus optimizing molecular structures and improving drug development success rates (Zhou et al., 2023).

Beyond drug discovery, XAI is also applied in fields such as protein engineering and medical image analysis. It not only aids researchers in understanding the reasoning behind AI model predictions but also provides decision-makers with the ability to trace model decisions. This is particularly important in high-risk applications such as drug approval and medical diagnosis, where the interpretability provided by XAI ensures regulatory compliance and reduces the risk of decision-making errors (Medina-Ortiz et al., 2025).

However, despite the significant potential of XAI, challenges remain, such as the trade-off between model interpretability and predictive performance, and the difficulty of achieving efficient interpretability for more complex tasks. Future research should focus on refining XAI techniques to enhance their applicability and practicality across various fields, thereby improving the trustworthiness and transparency of AI in high-risk domains.

### 6.2. Directions for model optimization and improvement

Model optimization and improvement of AI in protein-ligand prediction is an ongoing research area. Current computational methods, whether statistical energy functions or DL models, are essentially designed to fit the distribution patterns of certain features in the training data. Therefore, the primary task of model improvement is to enhance the quality of data, address class imbalances within datasets, and select appropriate methods for extracting effective features. In the construction of new datasets with relatively balanced classes, it is of the utmost importance to ensure the integrity and preservation of the original information. In order to expand the scale of datasets and enhance model robustness and generalization capabilities, techniques such as data augmentation or GANs can be employed to apply random transformations to original data or generate new samples. Furthermore, integration of traditional statistical energy functions with DL models allows for further optimization and enhancement. For instance, the design of more complex feature extraction methods that use domain knowledge or prior information to guide the feature extraction process is a potential avenue for further investigation. Alternatively, multi-modal fusion techniques, such as joint or multi-task learning, offer an alternative approach by integrating diverse data sources. This integration yields more comprehensive information, thereby enhancing a model's understanding of protein-ligand interactions, with particular relevance to PhPIs (Dhakal et al., 2021). Similarly, advanced DL architectures like GNNs are employed to better capture the complex features of these interactions.

Optimizing the model itself is also crucial. The ensemble learning method is utilized to combine predictions of multiple models to improve overall performance and robustness of the system. Transfer learning may be employed to share knowledge, parameters, or features between the source and target domains, effectively utilizing existing data and models. This approach accelerates the learning process for new tasks, reduces the necessity for extensive labeled data, and enhances model generalization and adaptability. In response to the black-box nature of DL models, research on how to build models that are highly interpretable allows researchers to comprehend the basis of model's predictions and contributes to a deeper understanding of PhPIs mechanisms. Furthermore, it is an interesting research direction to develop small sample learning and zero sample learning algorithms, so that the model can learn effectively even when there is only a little or no label data. As the gatekeeper of model construction, the scoring function of model evaluation plays a pivotal role in further adjusting parameters and optimizing models. A comprehensive evaluation of model performance allows us to understand how the model performs on different data sets, identify potential problems and bottlenecks, and provide guidance for improving the model. Therefore, the optimization of the scoring functions is also a promising direction for future model enhancement (Shen et al., 2020).

### 6.3. The Potential of DL in the Food Sector

The application of DL in the food industry is steadily advancing, offering robust tools and methodologies that have the potential to drive innovation and progress in the field. Polyphenols and proteins, as common biological compounds, are influenced by the quality, nutritional profile, and functionality of food products. Consequently, a more profound investigation into PhPIs is promising for uncovering novel scientific insights and infusing fresh momentum into research and applications within the food field.

DL techniques afford the opportunity to understand the mechanism behind PhPIs (Moon et al., 2024a). Leveraging DL algorithms for screening enables rapid identification of polyphenols exhibiting specific bioactivities and prediction of PhPIs. This process facilitates the discovery of complexes endowed with enhanced antioxidants, anti-inflammatory, and other beneficial biological properties, thereby driving advancements in functional food design and fostering innovation. Moreover, in the field of active substance delivery systems, DL technology expedites the screening of optimal protein-polyphenol combinations, enhancing the bioaccessibility of active ingredients while saving experimental time (Wang, Z. et al., 2024). What's more, researchers face stability issues when preparing emulsions using PhPIs, and DL offers more efficient solutions for experiments by directly predicting the binding affinity between the two, assessing the stability of the emulsion, and more (Wang et al., 2021). DL techniques play a pivotal role in investigation of altered allergenicity. PhPIs can influence protein structural conformation, rendering them more susceptible to degradation by digestive enzymes and consequently reducing exposure to allergenic epitopes, thereby mitigating the risk of food allergies (Zhao et al., 2023). Therefore, it is possible to predict the conformational effects of polyphenol incorporation on proteins, which can be rapidly screened for beneficial polyphenols and provide crucial support in mitigating the risk of food allergies.

The integration of DL into studies on PhPIs holds immense significance for the food industry. By investigating deeply into these interactions, it can offer theoretical support and scientific guidance for the development of healthier, more nutritious, and innovative food products. Thus, the adoption of DL technology in research on PhPIs will bring more innovation opportunities and development prospects for the food industry.

## 7. Conclusion

DL algorithms have been developed to predict PhPIs by constructing intricate neural network models, offering insights into binding modes and mechanisms. This computational approach establishes a critical foundation for optimizing functional properties, stabilizing emulsions, and enhancing bioactive compound delivery. By leveraging its powerful data-processing capabilities, DL streamlines the identification of interaction patterns, improving prediction accuracy and reliability through iterative model refinement.

Despite these advancements, challenges remain in the effective translation of DL-derived insights into practical applications. The complexity and heterogeneity of PhPIs usually demand extensive experimental validation, while the interpretability of DL models remains a bottleneck. Moreover, the reliance on high-quality, standardized data sets poses limitations in generalizability across diverse biological systems.

Future research should prioritize improving model architecture and training strategies to address these challenges. Integrating DL with complementary computational techniques, such as molecular docking and MD simulations, could enhance mechanistic understanding. Furthermore, advancements in XAI may bridge the gap between prediction and application by providing clearer rationales for binding site recognition and interaction specificity. As datasets expand and algorithms evolve, DL has the potential to drive the rational design of functional polyphenol-protein complexes, fostering innovation in food and biotechnology.

**Fundings**

The authors thank the National Natural Science Foundation of China (NSFC: 32202132), Young Elite Scientists Sponsorship Program by CAST (No. 2022QNRC001), the Priority Academic Program Development of Jiangsu Higher Education Institutions (PAPD) for financial support, and the Postdoctoral Fellowship Program of CPSF (GZB20230294).

**Conflict of interest**

The authors declared no known conflict of interest.